\documentclass[a4paper,fleqn,usenatbib,times]{mnras}

\usepackage{graphicx}	
\usepackage{amsmath}	
\usepackage{amssymb}	
\usepackage{mathtools}

\usepackage{xfrac}
\usepackage[capitalise]{cleveref}
     \crefname{equation}{Equation}{Equations}
     \crefname{figure}{Figure}{Figures}
     \crefname{table}{Table}{Tables}
     
\usepackage{bm}                                         
\usepackage{relsize}
\usepackage[table]{xcolor}
\usepackage{siunitx} 
\usepackage{booktabs}
\allowdisplaybreaks

\usepackage[T1]{fontenc}
\usepackage{ae,aecompl}

\usepackage{natbib}
\defcitealias{Laibe/Price/2014a}{LP14a}
\defcitealias{Laibe/Price/2014b}{LP14b}

\defcitealias{Hutchison/etal/2016b}{HLPM16b}

\title[Dust entrainment in photoevaporative winds]{On dust entrainment in photoevaporative winds}

\author[Hutchison, Price, Laibe, \& Maddison]{
Mark A. Hutchison,$^{1}$\thanks{E-mail: mhutchison@swin.edu.au}
Daniel J. Price,$^{2}$
Guillaume Laibe,$^{3}$
and Sarah T. Maddison$^{1}$
\\
$^{1}$Centre for Astrophysics \& Supercomputing, Swinburne University of Technology, Hawthorn, VIC 3122, Australia\\
$^{2}$Monash Centre for Astrophysics and School of Physics \& Astronomy, Monash University, Clayton, VIC 3800, Australia\\
$^{3}$School of Physics and Astronomy, University of St. Andrews, North Haugh, St. Andrews, Fife KY16 9SS, UK
}

\date{Accepted XXX. Received YYY; in original form ZZZ}

\pubyear{2016}

\begin{document}
\label{firstpage}
\pagerange{\pageref{firstpage}--\pageref{lastpage}}
\maketitle

\begin{abstract}
We investigate dust entrainment by photoevaporative winds in protoplanetary discs using dusty smoothed particle hydrodrodynamics (SPH). We use unequal-mass particles to resolve more than five orders of magnitude in disc/outflow density and a one-fluid formulation to efficiently simulate an equivalent magnitude range in drag stopping time. We find that only micron sized dust grains and smaller can be entrained in EUV driven winds. The maximum grain size is set by dust settling in the disc rather than aerodynamic drag in the wind. More generally, there is a linear relationship between the base flow density and the maximum entrainable grain size in the wind. A pileup of micron sized dust grains can occur in the upper atmosphere at critical radii in the disc as grains decouple from the low-density wind. Entrainment is a strong function of location in the disc, resulting in a size sorting of grains in the outflow---the largest grain being carried out between $10$--$20\,$AU. The peak dust density for each grain size occurs at the inner edge of its own entrainment region.
\end{abstract}




\section{Introduction}
\label{sec:introduction}

Small dust grains set the energy balance in the surface layers of protoplanetary discs. As a result, they determine the efficiency of photoevaporation \citep{Hollenbach/etal/1994}, a pressure-driven wind produced by high energy stellar radiation that heats and/or ionises gas located in the incident surface layers of the disc. Dust is a major source of opacity for impinging stellar radiation, but also an important link in collisional/chemical heating channels through which ultra-violet and X-ray radiation heat gas to escape velocities \citep[see][]{Tielens/Hollenbach/1985a,Glassgold/Najita/Igea/1997a,Alexander/Clarke/Pringle/2004b}. The role of dust in energy balance and photoevaporation is complicated by the aerodynamic drag that small dust grains feel in the presence of outflowing gas. Disc winds selectively entrain grains of different sizes at different radii resulting in dust populations that vary spatially \citep{Owen/Ercolano/Clarke/2011a} and temporally. The opacity and heating from entrained and semi-entrained dust grains create a complicated feedback mechanism that can only be resolved by combining radiative transfer and two-phase hydrodynamic simulations. Although \citet{Owen/etal/2010} found a method to couple the X-ray radiative transfer model of \citet{Ercolano/Clarke/Drake/2009} to \emph{gas} hydrodynamics, a self-consistent treatment of dust in hydrodynamic simulations of photoevaporation has not yet been attempted. Thus, the effect of an evolving dust phase in the disc and outflow (e.g. settling and aerodynamic drag) remains a source of uncertainty in many, if not all, photoevaporation models to date.

\begin{figure}
	\centering{\includegraphics[width=\columnwidth]{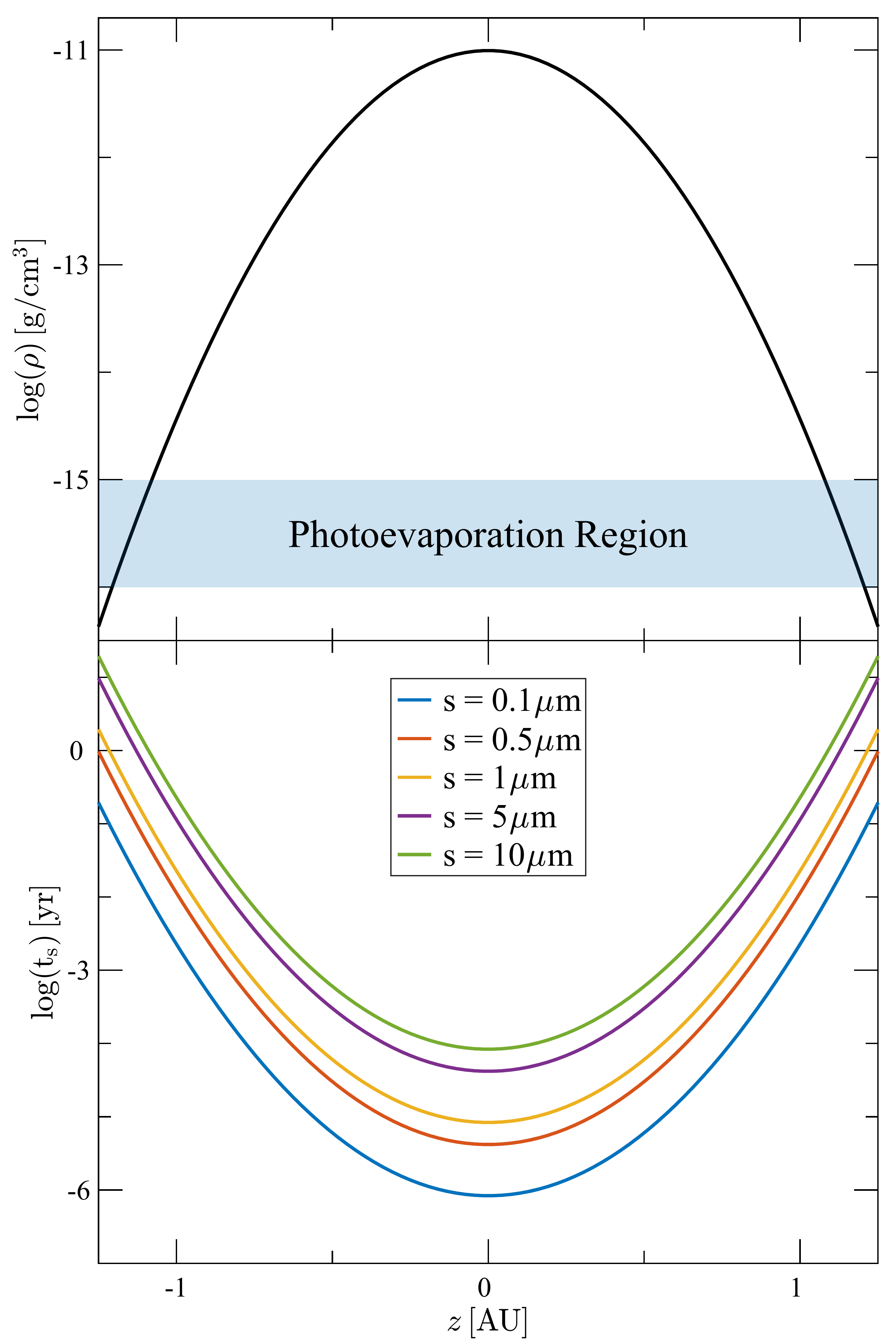}}
	\caption{Gas density (top) and drag stopping time (bottom) for different grain sizes, $s$, in a representative slice through a typical protoplanetary disc. The thermo-chemical models of \citet{Woitke/etal/2016} show that photoevaporation occurs high in the disc atmosphere at densities $\sim \! 4$--$5$ orders of magnitude below the local midplane density of the disc (shaded region). Dust entrainment during photoevaporation is numerically challenging due to the 5 orders of magnitude change in density (top) and 7 orders of magnitude change in stopping time (bottom) from disc midplane to atmosphere.}
	\label{fig:density_ts}
\end{figure}

The reason for this global oversight is that incorporating small dust grain dynamics into photoevaporation simulations is not trivial. Correctly accounting for aerodynamic drag in two-phase hydrodynamic simulations involving small dust grains has proven to be numerically difficult even for simple test cases, let alone disc atmospheres undergoing photoevaporation \citep{Laibe/Price/2012a,Laibe/Price/2012b}. Disc atmospheres cover a huge range in density and temperature---two quantities that strongly affect the drag timescale---and they vary in both the vertical and the radial directions. To illustrate the physical scales that need to be resolved, \cref{fig:density_ts} shows the gas density (top panel) and the drag stopping times for various sized dust grains (bottom panel) in a representative slice through a protoplanetary disc. What is numerically straightforward in one area of the disc may be very difficult in another. Furthermore, a single grain size initially distributed evenly throughout the disc will experience a mix of strong, intermediate, and weak drag regimes depending on local disc conditions, making it difficult to predict how a particular grain size should behave globally in the disc.

Fortunately, new techniques exist that allow us to model small dust grains much more efficiently and accurately than in the past \citep{Laibe/Price/2014a,Laibe/Price/2014b,Laibe/Price/2014c,Price/Laibe/2015,Loren-Aguilar/Bate/2014}. However, there are a few obstacles to overcome before we can directly apply these methods to photoevaporation. Foremost is the fact that these techniques have been developed using SPH \citep{Lucy/1977,Gingold/Monaghan/1977}. SPH concentrates resolution towards regions of high density, making it difficult to resolve a low-density outflow in the presence of a high-density disc midplane. Theoretically, this issue can be solved by using unequal-mass particles, but in practice there are a number of numerical complications that have prevented unequal masses from gaining much traction in the SPH community \citep[e.g. see][]{Rasio/Lombardi/1999,Monaghan/Price/2006}. Successfully circumventing these numerical obstacles is not straightforward; however, once achieved, the return is large. Not only do we gain access to small grain physics, but we also inherit SPH's intrinsic ability to handle free boundary conditions, making it perfect for next generation photoevaporation models simulating asymmetric features like planet-induced spiral density waves and/or warps.

We propose that self-consistent, two-phase hydrodynamic simulations of dusty photoevaporation will reveal new insights into how dust behaves in the upper atmospheres of photoevaporating discs. This added insight could have a significant impact on future radiative transfer calculations of dusty protoplanetary discs. In order to test this hypothesis, we first develop an unequal-mass, one-fluid SPH formalism that accurately and efficiently simulates two-phase hydrodynamics with large density ranges. Then, using a simple photoevaporation model, we self-consistently evolve gas and dust in the presence of photoevaporating flows. We find that entrained dust grains exhibit a rich variety of dynamic behaviours and properties in the wind that are subject to both disc and stellar conditions. Furthermore, we find that due to disc settling, the largest grain size in photoevaporative winds is not \emph{a priori} equal to the maximum grain size that can be supported by the wind. As a result, photoevaporative winds may be even less dusty than originally thought.

The paper is organised as follows: \cref{sec:photoevaporation} gives a brief overview of the different heating mechanisms responsible for photoevaporation; \cref{sec:sph_formalism} summarises the implicit, unequal-mass, one-fluid SPH formalism necessary to simulate both small dust grain physics and photoevaporation; \cref{sec:plane-parallel_atmosphere} introduces our disc model, explains why our SPH formalism is necessary, and demonstrates that it works; \cref{sec:benchmarking} benchmarks our code against a semi-analytic solution of dusty photoevaporation; and finally, using our new model, \cref{sec:dusty_photoevaporation} collates a suite of photoevaporation simulations that explore the behaviour of dust in various physical regimes. In a forthcoming paper, \citet[][hereafter, \citetalias{Hutchison/etal/2016b}]{Hutchison/etal/2016b}, we have developed a simpler, semi-analytic model that is able to recover many of the results found in this numerical study.


\section{Photoevaporation heating}
\label{sec:photoevaporation}

Photoevaporation comes in three flavours, generally classified by the wavelength or energy of the incident stellar flux on the disc: (i) ionising extreme-UV radiation (EUV; $13.6$--$100\,$eV), (ii) photodissociating far-UV radiation (FUV; $6$--$13.6\,$eV),  and (iii) X-ray radiation ($0.1$--$1\,$keV). Understanding which energy regime (or combination thereof) is predominately responsible for photoevaporation in protoplanetary discs is still an open question \citep[for a recent review, see][]{Alexander/etal/2014}. We briefly present an overview of the heating mechanisms for each type.

\subsection{EUV photoevaporation}
\label{sec:EUV}

The sharp cutoff between EUV and FUV radiation is marked by the ionisation of hydrogen---the most common element in protoplanetary discs. Consequently, EUV radiation has a particularly high absorption cross-section. This peaks at $13.6\,$eV and drops precipitously due to a rough $\nu^{-3}$ dependence, making it relatively independent of stellar spectrum \citep{Osterbrock/Ferland/2006}. The transition between neutral and ionised gas is abrupt and provides, to zeroth order, negligible heating to the bulk of the disc. As a result, gas and dust temperatures are coupled approximately up to the ionisation front. Beyond the ionisation front the gas is near isothermal ($T\simeq 10^4\,$K) with constant sound speed, $c_\text{s}\simeq 10\,$km/s. Taken together, these properties simplify EUV photoevaporation tremendously, allowing the flow to be determined by knowing only the EUV flux and base flow density \citep{Alexander/Clarke/Pringle/2006a}.

\subsection{FUV photoevaporation}
\label{sec:FUV}

Although FUV radiation is below the ionisation threshold for hydrogen, it is still energetic enough to drive significant heating in discs (as evidenced by FUV photodissociation regions in giant molecular clouds). However, FUV heating is considerably more complicated than EUV. For example, \citet{Tielens/Hollenbach/1985a} summarise at least 8 mechanisms by which FUV radiation is converted into gas heating. There are three crucial differences between EUV and FUV radiation: (i) attenuation of FUV flux in photoevaporating discs is dominated by dust and polycyclic aromatic hydrocarbon absorption \citep{Alexander/etal/2014}, thereby \emph{coupling} FUV photoevaporation to the dust evolution in the disc and \emph{decoupling} the dust temperature from that of the gas \citep[at low column density; see][]{Adams/etal/2004}; (ii) the launch point for the FUV flow is located in the atomic layer of the disc, well below the EUV ionisation front \citep{Gorti/Hollenbach/2009}; and (iii) the FUV mass-loss rate peaks at $\sim \! 5$--$10\,$AU, but the radially extended heating from FUV radiation actually makes it the dominant wind mechanism beyond $\sim\!100\,$AU \citep[see][]{Armitage/2011}.

\subsection{X-ray photoevaporation}
\label{sec:X-ray}

Like FUV radiation, X-rays have a complicated network of chemically dependent channels through which they can heat the disc. The relatively large spectral range and high energies cause X-ray absorption to preferentially occur in heavier elements via ionisation of inner (K-shell) electrons \citep{Glassgold/Najita/Igea/1997a}. However, hard X-rays ($\gtrsim1\,$keV) are too penetrating to be important for photoevaporation \citep{Ercolano/Clarke/Drake/2009}. The primary heating agent for the gas is not the ionisation of heavy elements themselves, but the subsequent release of photo/Auger electrons that collisionally ionise and/or heat the lighter atoms/molecules in the disc. Like FUV radiation, X-rays decouple the gas and dust temperatures in the disc's atmosphere. Furthermore, as much of the heavy elements are tied up in dust grains, X-ray heating is also sensitive to the location of the dust in the disc.\\

While the heating mechanism creating photoevaporation is different from one case to the other, the resultant outflow is hydrodynamically similar. This is critical because our goal in this paper to hydrodynamically characterise dust entrainment in photoevaporative winds, not to determine which flavour of photoevaporation is dominant in nature. Thus, as long as we can produce an outflow using just one of the methods, the techniques can be applied to others. Since EUV photoevaporation has the simplest heating mechanism and the fewest dust dependencies, we focus exclusively on EUV driven winds in this paper. 


\section{Numerical method}
\label{sec:sph_formalism}

Simulating small dust grains in a two-fluid code can be computationally challenging for four reasons \citep[see][]{Laibe/Price/2012a}. First, the explicit integration of the fluid equations requires an additional timestep restriction resulting in a timestep smaller than the minimum drag stopping time, i.e. $\Delta t < t_\text{s}$, in order to resolve the aerodynamic coupling between the gas and dust. Second, overdamping occurs between phases whenever the spatial criterion $h \lesssim c_\text{s}t_\text{s}$ is not met. Here $h$ is the resolution length of the simulation (particle smoothing length for SPH) and $c_\text{s}$ is the local sound speed of the gas. Third, each phase requires its own set of particles which approximately doubles memory requirements and simulation times. Fourth, two-fluid codes suffer from artificial clumping of dust below the resolution scale of the gas. 

The multi-phase, one-fluid formalism developed by \citet{Laibe/Price/2014a,Laibe/Price/2014b} overcomes all four of these issues, therefore we use this method to simulate photoevaporation in SPH. 

\subsection{One-fluid gas and dust mixture in SPH}
\label{sec:onefluid}

\subsubsection{Continuum equations}
\label{sec:continuum_equations}

\citet{Laibe/Price/2014a} showed that the fluid equations for a two-phase fluid of gas and dust can be re-formulated to describe a single fluid moving at the barycentric velocity of the mixture,
\begin{equation}
	\mathbf{v} \equiv \frac{\rho_\text{g} \mathbf{v}_\text{g}+\rho_\text{d} \mathbf{v}_\text{d}}{\rho}.
\end{equation}
Here, and throughout the rest of the paper, the subscripts g and d refer to gas and dust quantities, respectively. The total density $\rho = \rho_\text{g}+\rho_\text{d}$ can be rewritten in terms of the dust fraction $\epsilon \equiv \rho_\text{d}/\rho$ and the differential velocity between the two phases, $\Delta \mathbf{v} \equiv \mathbf{v}_\text{d} - \mathbf{v}_\text{g}$. Substituting these quantities into the fluid equations to remove any explicit dependance on individual phases, we are left with the following set of equations describing a single fluid
\begin{align}
	& \frac{\text{d}\rho}{\text{d}t} = - \rho (\nabla \cdot \mathbf{v}),
	\label{eq:onefluid_density}
\\
	& \frac{\text{d} \epsilon}{\text{d} t}  = -\frac{1}{\rho} \nabla \cdot \left[ \epsilon (1-\epsilon) \rho \Delta \mathbf{v} \right],
	\label{eq:onefluid_dustfrac}
\\
	&\frac{\text{d} \mathbf{v}}{\text{d} t}  = (1-\epsilon) \mathbf{f}_\text{g}
		 + \epsilon \mathbf{f}_\text{d} 
		 -\frac{1}{\rho} \nabla \cdot \left[ \epsilon (1-\epsilon) \rho \Delta \mathbf{v} \Delta \mathbf{v} \right] 
		 + \mathbf{f},
	\label{eq:onefluid_momentum}
\\
	& \frac{\text{d} \Delta \mathbf{v}}{\text{d} t} \! = \! -\frac{\Delta \mathbf{v}}{t_\text{s}}  
		+  (\mathbf{f}_\text{d} \! - \! \mathbf{f}_\text{g}) 
		 -  (\Delta \mathbf{v} \! \cdot \! \nabla) \mathbf{v}  
		 +  \frac{1}{2} \nabla \! \left[  (2\epsilon \! - \! 1) \Delta v^2  \right],
	\label{eq:onefluid_deltav}
\\
	&\frac{\text{d} u}{\text{d} t}  = - \frac{P_\text{g}}{(1-\epsilon) \rho} \nabla \cdot (\mathbf{v}
		 - \epsilon \Delta \mathbf{v}) 
		 + \epsilon (\Delta \mathbf{v} \cdot \nabla ) u 
		 + \epsilon \frac{\Delta v^2}{t_\text{s}},
	\label{eq:onefluid_energy}
\end{align}
where the convective derivative ($\text{d}/\text{d}t \equiv \partial/\partial t + \mathbf{v} \cdot\nabla$) now refers to a single fluid moving at the barycentric velocity. External forces exerted on the fluid are represented by $\mathbf{f}$, while the subscripts denote any force that applies to only a single phase, e.g. $\mathbf{f}_\text{g} = - \nabla P_\text{g}/\rho_\text{g}$. The gas pressure is represented by $P_\text{g}$ and $u$ is the internal energy. The stopping time is defined as
\begin{equation}
	t_\text{s} \equiv  \frac{\epsilon (1-\epsilon) \rho}{K},
\end{equation}
where $K$ is the drag coefficient which in general depends on the local gas and dust parameters. In this paper we will assume that $K$ is either constant or in the linear Epstein regime, suitable for small dust grains with low Mach numbers \citep[][also e.g. \citealt{Laibe/Price/2012b}]{Epstein/1924}. In the latter case, assuming spherical dust grains,
\begin{equation}
	K = \rho_\text{g}\rho_{d}\frac{4\pi}{3}\frac{s^2}{m_\text{grain}}\sqrt{\frac{8}{\pi \gamma}} c_\text{s},
	\label{eq:drag_constant}
\end{equation}
where $s$ is the grain size, $m_\text{grain}=\frac{4\pi}{3}\rho_\text{grain}s^3$ is the mass of each grain, and $\rho_\text{grain}$ is the intrinsic dust density. Thus the stopping time in the Epstein regime is equivalent to
\begin{equation}
	t_\text{s} = \frac{\rho_\text{grain}s}{\rho c_\text{s}} \sqrt{\frac{\pi \gamma}{8}}.
	\label{eq:stopping_time}
\end{equation}

\subsubsection{Discretised SPH equations}
\label{sec:discretised_sph_equations}

\citet[][hereafter referred to as \citetalias{Laibe/Price/2014b}]{Laibe/Price/2014b} transcribed \cref{eq:onefluid_density,eq:onefluid_dustfrac,eq:onefluid_momentum,eq:onefluid_deltav,eq:onefluid_energy} into a fully conservative set of SPH equations---including conservative shock-capturing terms. The discretised equations for a single set of particles (indexed hereafter using subscripts $a$, $b$, and $c$) that represent a mixture of both gas and dust are as follows:
\begin{align}
	& \rho_a = \sum_b m_b W_{ab}(h_a),
	\label{eq:SPH_density}
\\
	& \frac{\text{d} \epsilon_a}{\text{d} t}  = - \sum_b m_b \left[ \frac{ \epsilon_a (1-\epsilon_a)}{\chi^a_b \rho_a}  \Delta \mathbf{v}_a \cdot \nabla_a W_{ab}(h_a) \right.   \nonumber 
	\\	& \phantom{\frac{\text{d} \epsilon_a}{\text{d} t}  = - \sum_b m_b}  \left. {} 
		+ \frac{\epsilon_b (1-\epsilon_b)}{\chi^b_a \rho_b} \Delta \mathbf{v}_b \cdot \nabla_a W_{ab}(h_b) \right],
	\label{eq:SPH_dustfrac}
\\
	& \frac{\text{d} \mathbf{v}_a}{\text{d} t}  =  (1-\epsilon_a) \mathbf{f}_\text{g} + \mathbf{f}_a    \nonumber 
	\\	& \phantom{\frac{\text{d} \mathbf{v}_a}{\text{d} t}  =} {}
		- \sum_b m_b \left[ \frac{ \epsilon_a (1-\epsilon_a) \Delta \mathbf{v}_a}{\chi^a_b \rho_a}  \Delta \mathbf{v}_a \cdot \nabla_a W_{ab}(h_a) \right.   \nonumber 
	\\	&  \phantom{\frac{\text{d} \mathbf{v}_a}{\text{d} t}  = - \sum_b m_b} \left. {} + \frac{\epsilon_b (1-\epsilon_b) \Delta \mathbf{v}_b}{\chi^b_a \rho_b} \Delta \mathbf{v}_b \cdot \nabla_a W_{ab}(h_b) \right],
	\label{eq:SPH_momentum}
\\
	& \frac{\text{d} \Delta \mathbf{v}_a}{\text{d} t}  = - \frac{\Delta \mathbf{v}_a}{t_{\text{s},a}}
		 -  \mathbf{f}_\text{g}  +  \sum_b m_b \frac{\mathbf{v}_{ab}}{\chi^a_b \rho_a}  \Delta \mathbf{v}_a \cdot \nabla_a W_{ab}(h_a)    \nonumber 
	\\	& \phantom{\frac{\text{d} \Delta \mathbf{v}_a}{\text{d} t}  =} {}
		+ \frac{1}{2} \sum_b  \frac{m_b}{\chi^a_b  \rho_a}   \left[  (1 - 2 \epsilon_a) \Delta v_a^2     \right.  \nonumber
	\\	& \phantom{\frac{\text{d} \Delta \mathbf{v}_a}{\text{d} t}  = + \frac{1}{2} \sum_b  \frac{m_b}{\chi^a_b  \rho_a}} \left. {} 
		-   (1 - 2 \epsilon_b) \Delta v_b^2  \right]  \nabla_a W_{ab}(h_a)   \nonumber 
	\\	& \phantom{\frac{\text{d} \Delta \mathbf{v}_a}{\text{d} t}  =} {} + 
		\frac{1}{\epsilon_a (1 - \epsilon_a)} \sum_b  m_b   \left[ \frac{q_{\Delta \mathbf{v},a}^\text{AV}}{\chi^a_b \rho_a^2}  \nabla_a W_{ab}(h_a)  \right. \nonumber
	\\	& \phantom{\frac{\text{d} \Delta \mathbf{v}_a}{\text{d} t}  = + \frac{1}{\epsilon_a (1 - \epsilon_a)} \sum_b  m_b } \left. {} 
		+ \frac{q_{\Delta \mathbf{v},b}^\text{AV}}{\chi^b_a \rho_b^2}  \nabla_a W_{ab}(h_b)  \right], 
	\label{eq:SPH_deltav}
\\
	& \frac{\text{d} u_a}{\text{d} t} =  \epsilon_a \frac{\Delta v_a^2}{t_{\text{s},a}} 
		+ \sum_b m_b \frac{P_a + q_{ab,a}^\text{AV}}{\chi^a_b \rho_a \rho_{a}^\text{g}}  \mathbf{v}_{ab}^\text{g} \cdot \nabla_a W_{ab}(h_a)    \nonumber 
	\\	& \phantom{ \frac{\text{d} u_a}{\text{d} t} =}  {} - \sum_b m_b u_{ab} \frac{\epsilon_a}{\chi^a_b \rho_a} \Delta \mathbf{v}_a \cdot \nabla_a W_{ab}(h_a)    \nonumber 
	\\	& \phantom{ \frac{\text{d} u_a}{\text{d} t} =} {} + \frac{1}{1-\epsilon_a} \sum_b m_b \left[ \frac{Q_{ab,a}}{\chi^a_b \rho_a^2} F_{ab}(h_a) 
		+ \frac{Q_{ab,b}}{\chi^b_a \rho_b^2} F_{ab}(h_b)  \right. \nonumber
	\\	& \phantom{ \frac{\text{d} u_a}{\text{d} t} = + \frac{1}{1-\epsilon_a} \sum_b m_b} \left. {}
		- \frac{q_{\Delta \mathbf{v},a}^\text{AV}}{\chi^a_b \rho_a^2} \Delta \mathbf{v}_{ab} \cdot \mathbf{\hat{r}}_{ab} F_{ab}(h_a) \right],
	\label{eq:SPH_energy}
\end{align}
where,
\begin{align}
	& (1-\epsilon_a) \mathbf{f}_\text{g}   = - \sum_b m_b \left[ \frac{P_a + q_{ab,a}^\text{AV}}{\chi^a_b \rho_a^2} \nabla_a W_{ab}(h_a)  \right. \nonumber
	\\	& \phantom{(1-\epsilon_a) \mathbf{f}_\text{g}   = - \sum_b m_b } \left.  + \frac{P_b + q_{ab,b}^\text{AV}}{\chi^b_a \rho_b^2} \nabla_a W_{ab}(h_b)  \right].  
\end{align}
The artificial viscosity, differential velocity dissipation, and conductivity parameters are given by:
\begin{align}
	&  \makebox[0pt][l]{$q_{ab,a}^\text{AV} $}\phantom{Q_{ab,a} }  \equiv
	 	\begin{cases}
			-\frac{1}{2}(1-\epsilon_a) \rho_a	 v_{\text{sig},a} \mathbf{v}_{ab}^\text{g}\! \cdot \mathbf{\hat{r}}_{ab}, & \!\text{if } \mathbf{v}_{ab}^\text{g} \! \cdot \mathbf{\hat{r}}_{ab} \! < 0 
		\\
			0, & \!\text{if } \mathbf{v}_{ab}^\text{g} \! \cdot \mathbf{\hat{r}}_{ab} \!\geq 0,
		\end{cases}
\\	& \makebox[0pt][l]{$q_{\Delta \mathbf{v},a}^\text{AV} $}\phantom{Q_{ab,a} } \equiv \frac{1}{2} \epsilon_a (1-\epsilon_a) \rho_a v_{\text{sig},\Delta \mathbf{v}} \Delta \mathbf{v}_{ab} \cdot \mathbf{\hat{r}}_{ab},
	\label{eq:deltav_diss}
\\	& Q_{ab,a} \equiv \frac{1}{2} \alpha_u \rho_a v_{\text{sig},u} u_{ab},
\end{align}
with their corresponding signal speeds defined as:
\begin{align}
	\makebox[0pt][l]{$v_{\text{sig},a}$}\phantom{v_{\text{sig},\Delta \mathbf{v}}} & \equiv \alpha c_{\text{s},a} + \beta | \mathbf{v}_{ab}^\text{g} 
		\cdot \mathbf{\hat{r}}_{ab} |, & & \text{if } \mathbf{v}_{ab}^\text{g} \cdot \mathbf{\hat{r}}_{ab} < 0, 
\\[1ex]
	v_{\text{sig},\Delta \mathbf{v}} & \equiv \alpha_{\Delta \mathbf{v}} c_{\text{s},a},	
	\label{eq:deltav_sig_speed}
\\[1ex]
	& && \text{if } \mathbf{g} = 0  \nonumber \\[-2\jot]
	\makebox[0pt][l]{$v_{\text{sig},u}$}\phantom{v_{\text{sig},\Delta \mathbf{v}}} &\equiv \alpha_u 
		\smash{\left\{\begin{array}{@{}l@{}}
			\sqrt{\frac{\left| P_a - P_b \right|}{\overline{\rho}_{ab}}},
		   \\[2\jot]
			\left| \mathbf{v}_{ab} \cdot \mathbf{\hat{r}}_{ab} \right|,
		\end{array}\right.}  \\[-\jot] 
	& && \text{otherwise}. \nonumber
\end{align}
Here $\alpha$ and $\beta$ are the usual linear and quadratic SPH viscosity parameters, respectively, while $\alpha_{\Delta \mathbf{v}}$ and $\alpha_u$ are both dimensionless coefficients of order unity. We use $\mathbf{g}$ to represent the force of gravity. The choice of smoothing length, $h$, along with the associated $\nabla h$ correction terms, $\chi^a_b$, will be discussed in \cref{sec:settingh}. Likewise the choice and functional form of the smoothing kernel, $W_{ab}(h) \equiv W(|\mathbf{r}_a- \mathbf{r}_b|,h)$, will be discussed in \cref{sec:kernelchoice}. Note that we use the convention that double indices on unbarred quantities indicate a difference term (e.g. $\mathbf{v}_{ab} \equiv \mathbf{v}_a - \mathbf{v}_b$ and $u_{ab} \equiv u_a - u_b$), while barred quantities represent averages, e.g. $\overline{\rho}_{ab} \equiv \frac{1}{2} (\rho_a + \rho_b)$. Finally, $F_{ab}$ is defined such that $\nabla_a W_{ab} \equiv F_{ab} \mathbf{\hat{r}}_{ab}$.

The astute reader will notice that we have altered the equations from their original form in \citetalias{Laibe/Price/2014b}. First, we use an equivalent but alternate formalism for the artificial viscosity and thermal conduction terms following the method used in the \textsc{phantom} code \citep{Lodato/Price/2010,Price/Federrath/2010}, thereby avoiding the use of averaged quantities. Second, we dispense with the $\Delta \mathbf{v}$ dissipation switch in \cref{eq:deltav_diss} and set the corresponding signal speed equal to the sound speed, following the alternative dissipation formulation mentioned in \citetalias{Laibe/Price/2014b} which we found to produce better results. Third, we use a non-standard notation for the $\nabla h$ correction terms---including an additional index---in anticipation of employing unequal-mass particles. The extra index is needed when $h$ is a function of number density, as opposed to the more usual fluid density formulation (see \cref{sec:settingh}).

\subsection{Time-stepping}
\label{sec:time-stepping}

A key feature of using the one-fluid formalism is that each particle inherently contains the local information about both phases so there is no need to interpolate between them. This, coupled with the fact that the gas/dust drag equations can be solved exactly for simple drag coefficients \citep{Laibe/Price/2011}, means that implicit integration can be implemented in a straightforward manner. \citetalias{Laibe/Price/2014b} propose doing this in three steps via operator splitting: (i) rates for all particle quantities are computed explicitly \emph{without} the drag terms, (ii) differential velocity and drag-induced heating are computed using semi-analytic solutions and used to reconstruct explicit-like derivatives, and (iii) the particles' differential velocity and energy are updated in an explicit fashion using these derivatives. We adopt the explicit integration technique used by \citetalias{Laibe/Price/2014b} as well as their implicit integration method for the internal energy, but note that their integration technique for the differential velocity does not work as written in their paper. We therefore propose an alternative method for integrating $\Delta \mathbf{v}$ forward in time.

\subsubsection{Semi-analytic solutions for drag}
\label{sec:exact_solutions_for_drag}

Following \citetalias{Laibe/Price/2014b}, the change in internal energy due to drag over $\Delta t$ is calculated by integrating the final term in \cref{eq:onefluid_energy},
\begin{align}
	\Delta u_\text{drag} & = \frac{\epsilon}{t_\text{s}} \int_0^{\Delta t} \left[ \Delta \mathbf{v}^n \mathrm{e}^{-t'/t_\text{s}} + \mathbf{a}_0 t_\text{s} \left(1 - \mathrm{e}^{-t'/t_\text{s}}\right)  \right]^2 \text{d} t'   \nonumber
\\
	& =  \frac{\epsilon}{2} \left\{ 2 \mathbf{a}_0^2 t_\text{s} \Delta t - \mathrm{e}^{-2 \Delta t/t_\text{s}} \! \left( 1\!-\!\mathrm{e}^{\Delta t/t_\text{s}} \right) \! \left(\Delta \mathbf{v}^n \!- \! \mathbf{a}_0 t_\text{s}\right)   \right. \nonumber
\\
	& \phantom{=} \left. {} \cdot \left[ \Delta \mathbf{v}^n - \mathbf{a}_0 t_\text{s} + \mathrm{e}^{\Delta t/t_\text{s}} \left(\Delta \mathbf{v}^n + 3 \mathbf{a}_0 t_\text{s}\right)    \right] \right\}.
	\label{eq:implicit_energy_n+1}
\end{align}
We obtain an analytic equation for the differential velocity by assuming that the acceleration of $\Delta \mathbf{v}$ in a barycentric fluid due to all non-drag forces is independent of $t_\text{s}$ and approximately constant over each timestep. Then the relevant differential equation reduces to
\begin{equation}
	\frac{\text{d} \Delta \mathbf{v}}{\text{d} t} = - \frac{\Delta \mathbf{v}}{t_\text{s}} + \mathbf{a}_0,
\label{eq:drag_diffeq}
\end{equation} 
where $\mathbf{a}_0 \equiv (\text{d} \Delta \mathbf{v}/\text{d} t)^n_0$ and the subscript $0$ represents the contribution from all terms on the right-hand side of \cref{eq:SPH_deltav} that do not involve $t_\text{s}$. In this paper we are only interested in linear drag regimes, where $t_\text{s}$ is independent of $\Delta \mathbf{v}$, so the exact solution to \cref{eq:drag_diffeq} is
\begin{equation}
	\Delta \mathbf{v}(t) = \Delta \mathbf{v}(t_0) \mathrm{e}^{-t/t_\text{s}} + \mathbf{a}_0 t_\text{s} \left(1 - \mathrm{e}^{-t/t_\text{s}}\right),
\end{equation}
where $t_0$ is the time at the beginning of the timestep. Because this equation is only valid while $\mathbf{a}_0 \approx \text{constant}$ (i.e. over a single timestep), we find
\begin{equation}
	\Delta \mathbf{v}^{n+1} = \Delta \mathbf{v}^n \mathrm{e}^{-\Delta t/t_\text{s}} + \mathbf{a}_0 t_\text{s} \left(1 - \mathrm{e}^{-\Delta t/t_\text{s}}\right).
	\label{eq:implicit_deltav_n+1}
\end{equation}

\subsubsection{Implicit integration}
\label{sec:implicit_integration}

There is an important difference between \cref{eq:implicit_energy_n+1} and \cref{eq:implicit_deltav_n+1} which ultimately affects how we are able to update $u$ and $\Delta \mathbf{v}$. Because $\Delta u_\text{drag}$ exclusively accounts for all drag heating and can be clearly separated from all other heating processes, we are free to update $u$ using the operator splitting techniques described in \citetalias{Laibe/Price/2014b}, modulo a missing factor of $1/2$ in equation (97) of their paper. However, the drag contribution to $\Delta \mathbf{v}^{n+1}$ is inseparably dependent on its non-drag constituents and must be treated differently so as to not double count the contribution from $\mathbf{a}_0$.

The key point in updating $\Delta \mathbf{v}$ is that we \emph{already} have the semi-analytic solution in \cref{eq:implicit_deltav_n+1} determining what $\Delta \mathbf{v}^{n+1}$ should be at the end of the timestep. The only real source of error in using this equation comes from assuming $\mathbf{a}_0$ is constant over $\Delta t$---an assumption we already make when using our explicit timestep. Thus we propose initially calculating $\Delta \mathbf{v}$ directly using \cref{eq:implicit_deltav_n+1} in the predictor step, then re-calculating $\Delta \mathbf{v}$ using updated values for $t_\text{s}$ and $\mathbf{a}_0$, and averaging the two results in the corrector step. In equation form, this procedure can be written as follows: 
\begin{align}
	&\Delta \mathbf{v}^{**} =  \Delta \mathbf{v}^n \mathrm{e}^{-\Delta t/t_\text{s}} + \mathbf{a}_0 t_\text{s} \left(1 - \mathrm{e}^{-\Delta t/t_\text{s}}\right), &  \text{(predictor)}
\\
	& \negthickspace
	\begin{rcases}
		\Delta \mathbf{v}^{*} = \Delta \mathbf{v}^n \mathrm{e}^{- \Delta t/t_\text{s}^*} + \mathbf{a}_0^* t_\text{s}^* \left(1 - \mathrm{e}^{-\Delta t/t_\text{s}^*}\right),
		\\
		\Delta \mathbf{v}^{n+1} = \frac{1}{2} \left( \Delta \mathbf{v}^{**} + \Delta \mathbf{v}^{*} \right).
	\end{rcases} & \text{(corrector)}
\end{align}
As a final note, evolving $\Delta \mathbf{v}$ only makes sense while $\epsilon > 0$. Since it is perfectly acceptable to have particles with zero dust fraction, we forcibly set $\frac{\text{d} \Delta \mathbf{v}}{\text{d} t} = 0$ whenever $\epsilon = 0$. Failing to do so often results in unbounded growth in $\Delta \mathbf{v}$.

\subsection{Unequal-mass particles}
\label{sec:unequalmassparticles}

Equal-mass particles in SPH naturally concentrate in regions of high density, making it difficult to resolve low-density phenomena surrounding high-density structures. In principle, concentrating particles in the densest objects automatically fulfils the role that complicated adaptive mesh refinement schemes play in grid-based codes which adjust resolution on the fly. The caveat is that SPH almost always has a fixed number of particles so resolving high density regions comes at the expense of poorly resolving low density regions. Hence the only way to increase resolution in low density regions is to add more particles. This, however, is inefficient and results in severely curtailed timesteps due to over-resolved structures. Therefore, SPH with equal-mass particles can only be applied to photoevaporation problems where the relevant density range is $\lesssim3$ orders of magnitude.

Allowing unequal masses provides a potential fix to this problem by adding another degree of freedom to the density. The result is that the same density profile can be represented by multiple particle configurations by adjusting the mass of the particles. This degeneracy allows us to redistribute the particles in a more convenient fashion and, ultimately, extend the profile to lower densities. Furthermore, it conveniently allows us to sidestep the issue of very small timesteps. The difficulty in using unequal masses is that it falls on us to tailor the particle-mass configuration for every simulation in order to ensure that the physically relevant processes are adequately resolved. At the moment, this is more of an art than a science and may not be straightforward, or even possible, in all cases.

\subsubsection{Setting h}
\label{sec:settingh}

Traditionally, the smoothing length for equal-mass particles is set using:
\begin{equation}
	h_a = \eta \left( \frac{m_a}{\rho_a} \right)^{1/\nu},
	\label{eq:varh_equal_smoothing}
\end{equation}
where $\nu$ is the number of spatial dimensions and $\eta$ is a dimensionless factor of order unity that controls the number of neighbours per particle. \Cref{eq:SPH_density,eq:varh_equal_smoothing} form a set of non-linear equations that must be solved at each timestep to ensure self-consistency \citep{Monaghan/2002,Price/Monaghan/2007}. For equal-mass particles, the above method produces a smoothing length that is a function of the particle coordinates, i.e. $h^\nu \rho \sim \text{constant}$. However, the mass dependence in \cref{eq:varh_equal_smoothing} destroys this coordinate relation when using unequal-mass particles. To remove the mass dependence and regain this relationship, we express $h$ in terms of number density, $n$, as defined by \citet{Koshizuka/Nobe/Oka/1998} and later adapted for SPH by \citet{Hu/Adams/2006}. Then the set of equations governing $h$ for unequal-mass particles is \citep{Price/2012}
\begin{align}
	&n_a = \sum_b W_{ab}(h_a),
	\label{eq:varh_unequal_numdens}
\\
	&h_a = \eta \left( \frac{1}{n_a} \right)^{1/\nu}.
	\label{eq:varh_unequal_smoothing}
\end{align}
Of course the density is still required by the hydrodynamic equations and is computed via \cref{eq:SPH_density}, so $n$ can be computed alongside $\rho$ at no extra computational cost.

Changing the functional form of $h$, on the other hand, requires a different formulation for the $\nabla h$ correction terms that usually appear in \cref{eq:SPH_dustfrac,eq:SPH_momentum,eq:SPH_deltav,eq:SPH_energy}. These terms can be obtained by re-deriving the fluid equations from the Lagrangian and carefully accounting for the new definition of $h$ in \cref{eq:varh_unequal_numdens,eq:varh_unequal_smoothing}. These terms were first derived by D.J. Price and later reported by \citet{Merlin/etal/2010}:
\begin{equation}
	\chi^a_b \equiv \left[ 1 + \frac{\zeta_a/m_b}{\Omega_a^*}  \right]^{-1},
\end{equation}
where
\begin{equation}
	\zeta_a \equiv \frac{\partial h_a}{\partial n_a}  \sum_c m_c \frac{\partial W_{ac}(h_a)}{\partial h_a},
\end{equation}
and
\begin{equation}
	\Omega_a^* \equiv 1- \frac{\partial h_a}{\partial n_a}  \sum_c \frac{\partial W_{ac}(h_a)}{\partial h_a}.
\end{equation}
We use the notation $\Omega^*$ to avoid confusion with the usual equal-mass $\nabla h$ correction term and the summation index $\text{c}$ to emphasis that $\zeta$ and $\Omega^*$ require their own summation over the particles (again performed alongside $\rho$). It is straightforward to show that, apart from the alternate correction terms, the form of the one-fluid equations remain unchanged when evolving $h$ using a number density formulation.

\subsubsection{Kernel choice}
\label{sec:kernelchoice}

All equal-mass tests performed in this paper use the M6 and double-hump M6 quintic spline kernel as prescribed by \citet{Laibe/Price/2012a}. We found out that the M6 kernel is unsuitable for use with unequal-mass particles because it is susceptible to the pairing instability. The necessary condition that prevents particle pairing from occurring is a kernel with a non-negative Fourier transform \citep{Dehnen/Aly/2012}. The Wendland kernels have this as one of their defining properties \citep{Wendland/1995}, making them ideal candidates for unequal-mass simulations. We use the Wendland $C^2$ kernel for all unequal-mass simulations. This decision was purely based on computational cost (proper analysis of the performance and accuracy of the different kernels using unequal masses is beyond the scope of this paper).

Kernels are usually expressed as a generic function of the smoothing length $h$ and a dimensionless distance $q$ between particle pairs,
\begin{equation}
	W(\mathbf{r},h) \equiv \frac{\sigma}{h^\nu}f(q), \qquad \text{where } q \equiv \frac{|\mathbf{r}_{ab}|}{h_a}.
\end{equation}
The particular form of $f(q)$ for the Wendland $C^2$ is
\begin{equation}
	f(q) = \begin{cases}
		\left( 1-\frac{q}{2} \right)^3_+ \left( 1+\frac{3q}{2} \right), & \qquad \text{if } \nu = 1 
	\\[2ex]
		\left( 1-\frac{q}{2} \right)^4_+ (1+2q), & \qquad \text{if } \nu = 2,\, 3,
	\end{cases}
\end{equation}
where we define $(\cdot)_+ \equiv \text{max} \{ 0,\cdot \} $. The relevant normalisation factors are $\sigma = \left[\frac{5}{8}, \frac{7}{4\pi}, \frac{21}{16\pi}\right]$ in $[1,2,3]$ dimensions and the resolution length factor we use throughout is $\eta = 1.3$. Note the value of $\eta$ can dramatically affect stability when using the Wendland kernels \citep[for a discussion about choosing an appropriate $\eta$, see][]{Dehnen/Aly/2012}.


\section{Plane-parallel atmosphere}
\label{sec:plane-parallel_atmosphere}

Before proceeding further, we wish to demonstrate that the algorithms above are not only necessary, but sufficient for SPH to be used as a viable tool for simulating the range of densities and stopping times shown in \cref{fig:density_ts}. Consider a thin, plane-parallel slab of gas and dust placed in non-rotating, vertical, hydrostatic equilibrium created from pressure-gravity balance. Using the vertical component of gravity from a central star with mass $M$ and located a distance $R$ from the midplane of the slab, i.e.,
\begin{equation}
	\mathbf{g} = \frac{\mathcal{G} M z}{\left(R^2+z^2\right)^{3/2}}\mathbf{\hat{z}},
	\label{eq:vertical_gravity}
\end{equation}
we obtain a disc-like atmosphere with physical densities and temperatures. In particular, for an isothermal disc in near vertical hydrostatic equilibrium, the density can be expressed as,
\begin{equation}
	\rho(z) = \rho_\text{g,0} \, \mathrm{exp} \left[-\frac{z^2}{2H^2}\right],
	\label{eq:isothermal_density_profile}
\end{equation}
where $\rho_\text{g,0}$ is the midplane density and $H$ is the scale height of the disc \citep[e.g.][]{Laibe/Gonzalez/Maddison/2012}.

\begin{figure}
	\centering{\includegraphics[width=\columnwidth]{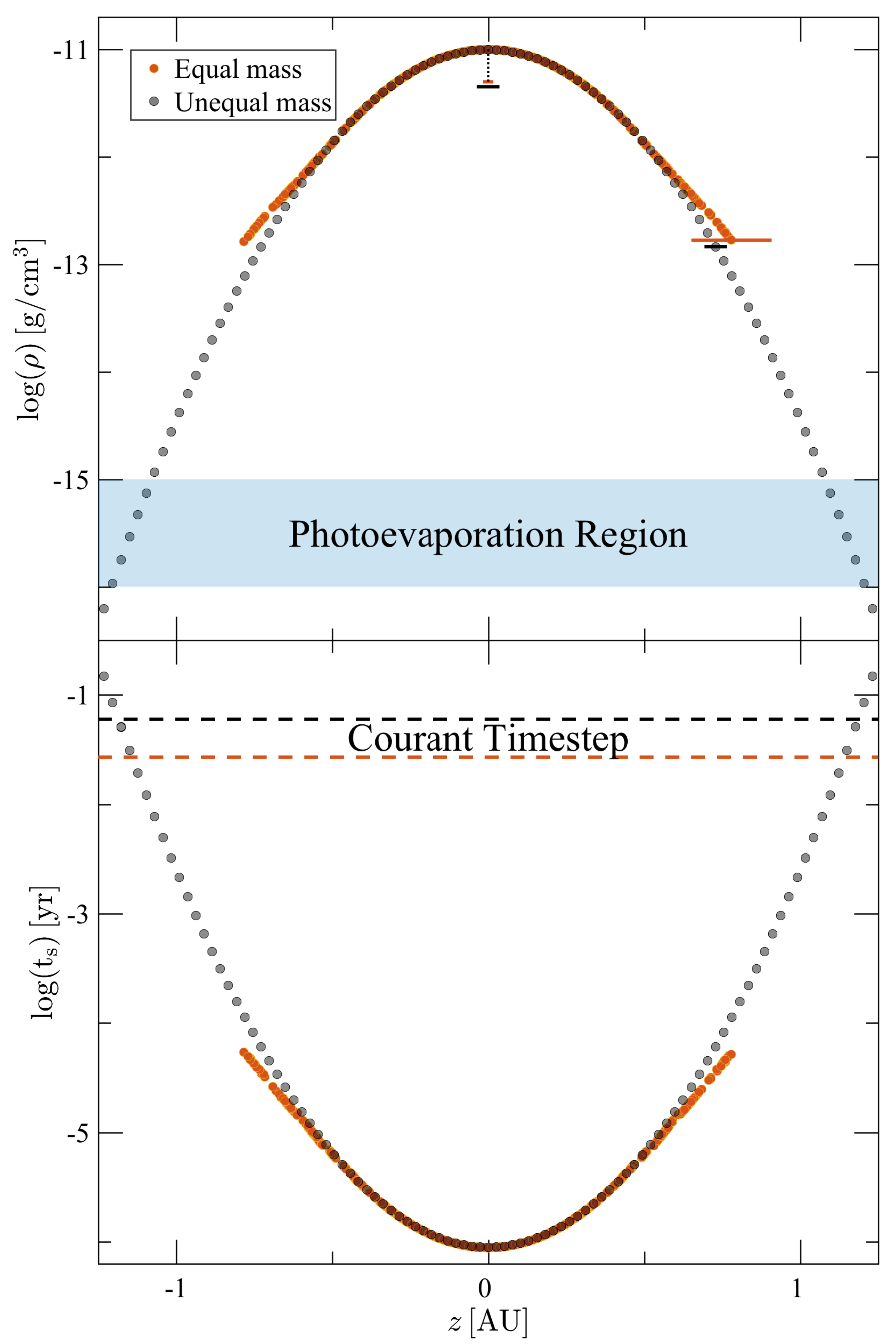}}
	\caption{Gas density (top) and stopping time (bottom) in two numerical simulations using equal-mass particles (brown) and unequal-mass particles (grey). The solid (dashed) brown and black lines in the top (bottom) panel indicate the smoothing length (Courant timestep) for each mass type. Equal-mass particles are unsuitable for simulating photoevaporation in SPH due to poor resolution in the surface layers of the disc (top), the inability to reach sufficiently low densities (top), and timestep restrictions from a crowded midplane (bottom). Having the minimum stopping time so much smaller than the Courant timestep suggests the need for an implicit timestepping algorithm. Our use of unequal-mass particles resolves the density range (top) while the one-fluid formalism with implicit timestepping handles the range in stopping time (bottom).}
	\label{fig:compare_equal_unequal_properties}
\end{figure}

This plane-parallel atmosphere contains all of the physics needed to calculate the drag properties shown in \cref{fig:density_ts}. The density profile from the top panel can be reproduced by choosing the following physical parameters: $M=1\,M_{\odot}$, $\rho_\text{g,0} = 10^{-11}\,\text{g/cm}^3$, $R=5\,$AU and $H=0.25\,$AU. Using these parameters, \cref{fig:compare_equal_unequal_properties} contrasts two identical discs in hydrostatic equilibrium created from equal-mass particles (brown) and unequal-mass particles (grey). To aid visualisation, only $5050$ particles are used in each case. The top panel overlays the density profiles for both discs along with the shaded photoevaporation region from \cref{fig:density_ts}. Importantly, equal-mass particles cannot resolve the photoevaporation region simultaneously with the disc midplane. Furthermore, the lack of resolution at low densities causes inaccuracies, namely the flared edges in the density profile. Increasing the number of particles is not a solution; even with over a million particles, we only secure $\sim \!10$ particles to resolve the photoevaporation region. 

Even if it were possible to resolve the density with equal-mass particles, the resulting restriction on the timestep from using equal-mass particles would be prohibitively small. The dashed lines in the bottom panel of \cref{fig:compare_equal_unequal_properties} show that the Courant timestep for equal-mass particles is already a factor of $\sim \! 2$ smaller than the unequal-mass particles. By the time one did add enough equal-mass particles to minimally resolve the photoevaporation region, the Courant timestep would be smaller than the drag stopping time. Thus, using equal-mass particles to achieve high resolutions in the photoevaporation region while maintaining a reasonable timestep is impractical.

However, using unequal-mass particles does not solve everything. In addition to the Courant timestep, the bottom panel also shows the drag stopping time for $s=0.1\,\mu$m dust grains, our smallest and most restrictive grain size in this study. The $\sim \! 5$ orders of magnitude difference between the Courant and drag limited timesteps shows how difficult it is to simulate small dust grains in discs explicitly. Our implicit scheme---made possible by the one-fluid approach---overcomes this issue and allows us to default to the Courant timestep. Furthermore, the one-fluid formalism allows us to accurately simulate the entire range of drag regimes exhibited by $0.1\,\mu$m grains or any other grain size. The accuracy of our implicit timestepping scheme and one-fluid formalism is tested hereafter; nevertheless, \cref{fig:compare_equal_unequal_properties} demonstrably shows that unequal-mass particles, working in tandem with the one-fluid formalism and implicit timestepping, are capable of simulating the physical conditions required by dusty photoevaporative winds. To our knowledge, it is currently the \emph{only} numerical method capable of doing so.

\section{Numerical tests}
\label{sec:benchmarking}

We have extensively benchmarked our code \textsc{gdphoto} on standard dust/gas test problems from the literature: \textsc{dustybox}, \textsc{dustywave}, \textsc{dustyshock}, and \textsc{dustydisc} \citep{Laibe/Price/2011,Laibe/Price/2012a,Price/Laibe/2015}. We catalogue a subset of these results---particularly those involving unequal-mass particles and implicit timestepping---in \cref{sec:standard_tests}. The only remaining segment of code to benchmark is our thermal energy switch used to create photoevaporation. We do this using the analytic plane-parallel wind solution and associated test problem proposed by \citet{Hutchison/Laibe/2016}, enhanced with a semi-analytic dust phase from \citetalias{Hutchison/etal/2016b} (in order to better suit the needs of our two-phase code). Following the \textsc{dusty-} naming convention from \citet{Laibe/Price/2012a}, we name this test \textsc{dustyphoto}.

\subsection{\sc{dustyphoto}}
\label{sec:photoevaporation_test}


The \textsc{dustyphoto} test uses the plane-parallel atmosphere outlined in \cref{sec:plane-parallel_atmosphere} because its simplified geometry allows an analytic treatment of photoevaporation while still accurately describing the vertical flow from discs on a local scale (see \cref{sec:plane-parallel_approximation}). Because the two-phase disc dynamics are verified in the \textsc{dustydisc} test in \cref{sec:dustydisc}, the \textsc{dustyphoto} test specifically measures the accuracy of the thermal energy switch responsible for photoevaporation (described below).

\subsubsection{Setup}
\label{sec:dustyphoto_setup}

The physical parameters for the disc are $M=1\,M_{\odot}$, $\rho_\text{g,0} = 10^{-11}\,\text{g/cm}^3$, $R=5\,$AU and $H=0.25\,$AU. We create the disc by placing $200\,028$ particles on a uniform (staggered) lattice inside a Cartesian box, $(x,z) \in [-6H , 6H]$, with periodic horizontal boundary conditions and dynamic vertical boundaries, as proposed by \citet{Hutchison/Laibe/2016}. The dynamic boundaries are created by converting the first ionised particles at $t=0$ into boundary particles and constraining them to move strictly in the vertical direction at the local wind speed prescribed by the analytic solution:
\begin{equation}
	v_\text{g} = c_\text{s} \sqrt{-\mathrm{W}_0 \! \! \left[ -\exp{ \left( -\frac{2\mathcal{G}M}{c_\text{s}^2 \sqrt{R^2+z^2}} - 1 \right)  }   \right]},
	\label{eq:plane-parallel_solution}
\end{equation}
where $\text{W}_0$ is the main branch of the Lambert $\text{W}$ function \citep{Corless/etal/1996}. Because the velocities are $z$ dependent, the boundary particles will slowly drift apart over time. Thus, it is important to initialise the run with enough boundary particles to prevent simulation particles from squeezing through into empty space. To be safe we employ $\sim \!20\,000$ boundary particles on each side of the disc.

The dust phase is comprised of a single grain size $s=2\,\mu$m, chosen to be near the maximum entrainable grain size, as this will provide the most stringent test of our drag coupling in the flow. In order to be consistent with the assumptions of the semi-analytic model from \citetalias{Hutchison/etal/2016b}, we enforce $\epsilon = 0.01$ for all neutral particles during the simulation, i.e. disc settling is not allowed to take place. We assume each grain has an intrinsic dust density $\rho_\text{grain} = 3\,$g/cm$^3$, while the SPH particle mass is set using the iterative method described in \cref{sec:dustydisc}. The solution for the dust is obtained by numerically integrating the following ordinary differential equation,
\begin{equation}
	v_\text{d} \frac{\text{d} v_\text{d}}{\text{d}z} = \frac{K}{\rho_\text{d}} \left( v_\text{g} - v_\text{d} \right) - \frac{\mathcal{G} M z}{\left( R^2 + z^2 \right)^{3/2}},
	\label{eq:dust_ode_with_dim}
\end{equation}
with the initial condition $v_\text{d,i}=0$. The gas and dust densities are obtained from the relation $\dot{m}=\rho v =  \rho_\text{i} v_\text{i}$  where $\dot{m}$ is a constant and the subscript $\text{i}$ on $\rho$ and $v$ denotes initial wind values at the ionisation front.

\subsubsection{Thermal energy switch}
\label{sec:dustyphoto_thermal_energy_switch}

Photoevaporation is created using an EUV thermal energy switch similar to \citet{Alexander/Clarke/Pringle/2006a}. We instantaneously heat any fluid that falls below the ionisation front density $\rho_\text{i}$, empirically selected from the photoevaporation region identified in \cref{fig:density_ts}. This empirical condition is necessary because we cannot integrate column densities along lines of sight in a plane-parallel atmosphere. For convenience, we parameterise $\rho_\text{i}$ using its own relative fraction to the midplane density, $\xi \equiv \frac{\rho_\text{i}}{\rho_\text{g,0}}$. Effectively, this parameterisation allows us to separately adjust the strength of the two physical quantities primarily responsible for the location of the ionisation front in the disc---the radial density profile (via $\rho_\text{g,0}$) and the EUV ionisation flux (via $\xi$).

The ionised gas is assumed to be isothermal with a temperature of $T=10^4\,$K and a sound speed $c_\text{s}=\sqrt{k_\text{B}T/\mu}$. Here $k_\text{B}$ is Boltzmann's constant and $\mu$ is the mean molecular weight---taken to be $0.88\,$u for ionised gas. The benefit of setting the ionisation front using the density is that it self adjusts the physical location according to the dynamics within the disc. This attribute is important for two reasons: First, the onset of photoevaporation creates large perturbations in the disc, particularly compression and expansion oscillations at the surface. Second, without accretion to resupply our disc with material, we never reach a true dynamic equilibrium in our disc. In either case, we need to be able to determine the location of the ionisation front without adding artefacts.

Although not strictly necessary, we find that we can converge to the analytic solution more quickly if we gradually raise the ionisation temperature from the isothermal disc temperature at $t=0$ to $10^4\,$K over a period of $\sim \!1.5$ years. Unlike the \textsc{dustydisc} test, there is no need to relax the disc prior to starting simulations. We start our disc in isothermal equilibrium, consistent with the initial density profile in \cref{eq:isothermal_density_profile}, but immediately default to an adiabatic equation of state whenever the gas particles are neutral, i.e. $P = (\gamma-1) \rho u$ where $\gamma = 5/3$. This allows us to capture the compressional heating caused by the ionised wind and to resolve the otherwise discontinuous temperature difference between the neutral disc and the ionised wind. The sound speed for neutral particles is calculated according to $c_\text{s}=\sqrt{\gamma k_\text{B} T/\mu}$, where $\mu = 2.34\,$u.

\subsubsection{Solution}
\label{sec:dustyphoto_solution}

\begin{figure*}
	\centering{\includegraphics[width=\textwidth]{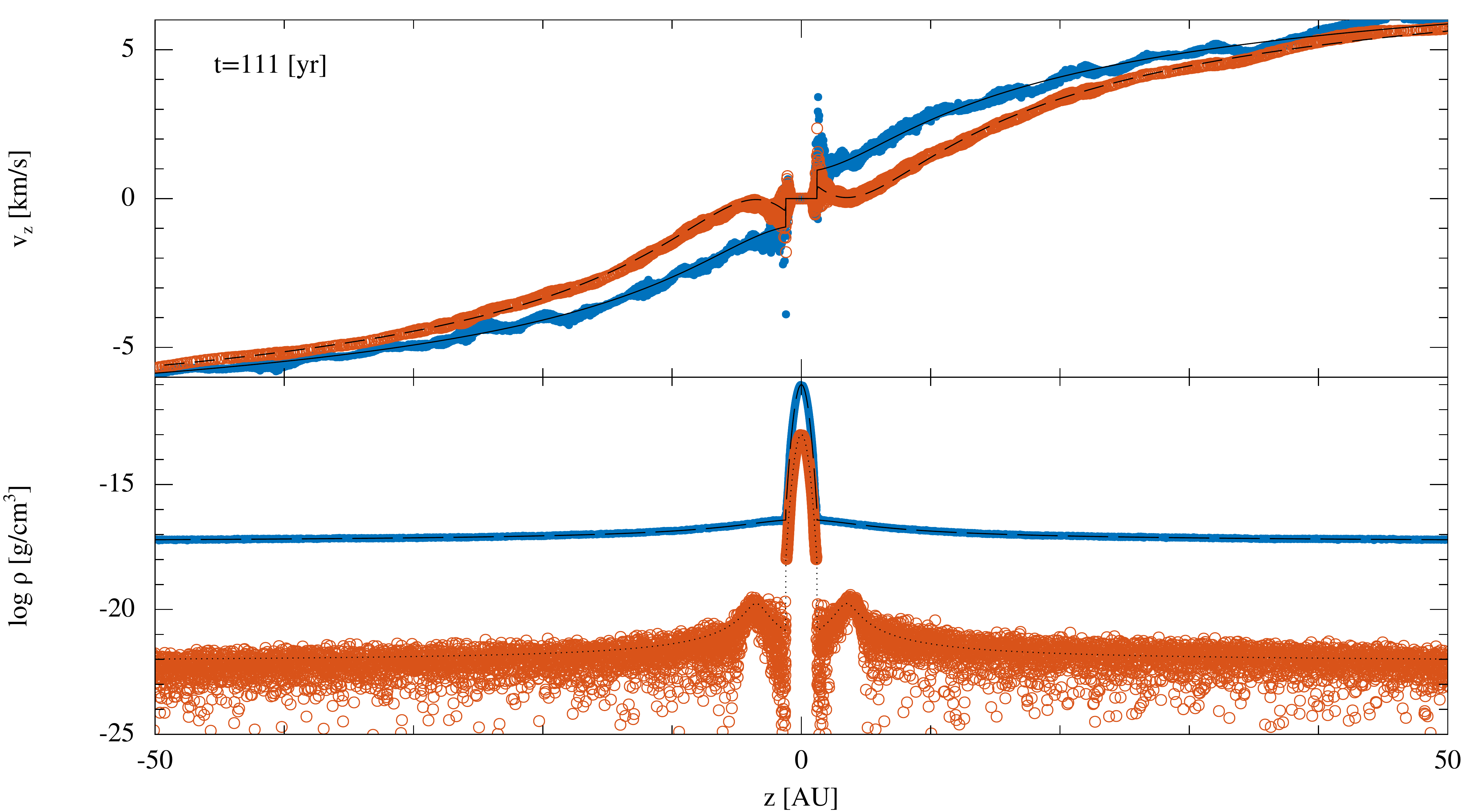}}
	\caption{Velocity (top) and density (bottom) profiles in the \textsc{dustyphoto} test after $111\,\text{yr} \approx 10\,$orbits using $200\,028$ SPH particles. The numerical solution for the gas (solid blue dots) and dust (open brown circles) phases are plotted together with the semi-analytic model (black solid, dashed, and dotted lines) from \citetalias{Hutchison/etal/2016b}. Using the dynamic boundary conditions proposed by \citet{Hutchison/Laibe/2016}, the gas converges to the analytic solution almost immediately. The dust velocity also converges very quickly to its semi-analytic solution, but the dust density takes much longer to reach its steady state, which lags behind the moving boundary by $\sim \! 50\,$AU.}
	\label{fig:analytic_photo_test}
\end{figure*}

\Cref{fig:analytic_photo_test} shows the velocity and density profiles of the gas and dust after $111\,$ years ($\approx 10\,$orbits) plotted against their analytic and semi-analytic solutions, respectively. The stationary shock created by the ionisation front causes nearly discontinuous jumps/drops in the density, velocity, pressure, and internal energy (cf. \cref{fig:shock_test_plot}). In order to match the semi-analytic model to our numerical solution, we visually fit the initial gas and dust density in the wind using $\dot{m}$. The values used in \cref{fig:analytic_photo_test} are $\dot{m}_\text{g} \approx 3.6 \times 10^{-12}\,\text{g}\, \text{cm}^{-2}\text{s}^{-1}$ and $\dot{m}_\text{d} \approx 5.8 \times 10^{-15}\,\text{g}\, \text{cm}^{-2}\text{s}^{-1}$. The $L_2$ errors computed by \textsc{splash} \citep{Price/2007} are in every case less than $1\%$, suggesting that our photoevaporation mechanism is robust and working as expected. 

Note while running this test, we observe a numerical instability in the one-fluid SPH equations that causes $\epsilon$ and $\Delta \mathbf{v}$ to diverge at the base of the outflowing wind. We observe a similar phenomenon while running the \textsc{dustydisc} test in \cref{sec:dustydisc}. We circumvent this instability by using the non-conservative SPH formulation prescribed in \cref{sec:numerical_instability_method_1}.


\begin{figure*}
	\centering{\includegraphics[width=\textwidth]{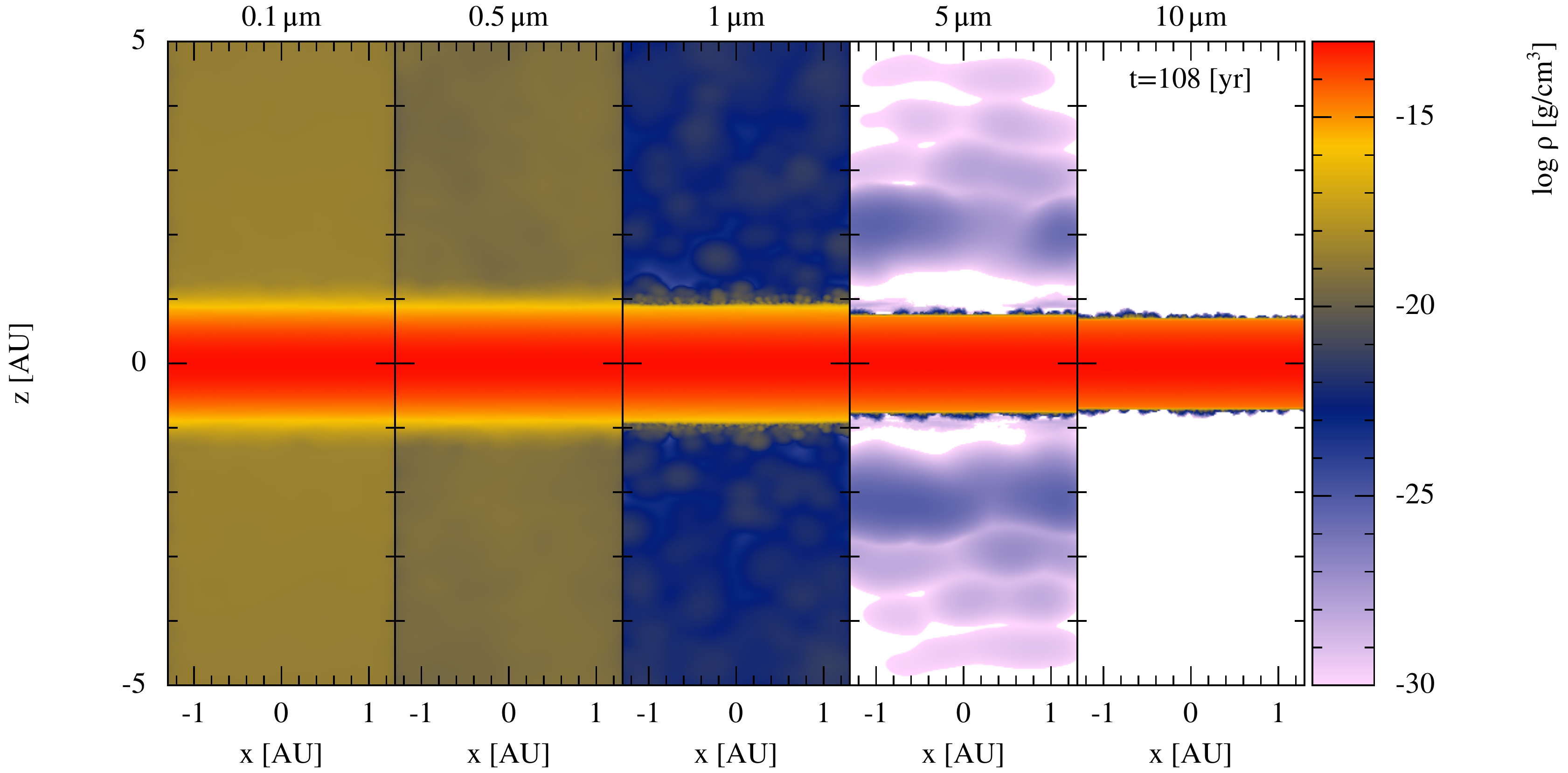}}
	\caption{Dust densities for five different grain sizes in a photoevaporating vertical disc slab using $200\,028$ particles. For comparison, each grain size is identically distributed through the disc with a dust-to-gas ratio of $0.01$. Only grains $\lesssim1\,\mu$m are entrained in the outflowing winds. The decaying density oscillations the appear in the $5\,\mu$m panel are numerical artefacts similar to the post-shock oscillations observed in the \textsc{dustyshock} problem in \cref{sec:dustyshock}.}
	\label{fig:grainsize}
\end{figure*}

\section{Dusty photoevaporation}
\label{sec:dusty_photoevaporation}

Having shown that \textsc{gdphoto} can accurately simulate the gas and dust dynamics in photoevaporating discs, we are now ready to allow dust settling within the disc during photoevaporation. At the same time, we examine how grain size, base flow density, and radius affect dust entrainment in the wind. Then we discuss the validity of the non-rotating, plane-parallel approximation, and finally, we look for any residual effects in the gas produced by the back reaction of entrained dust grains. Except where otherwise indicated, we continue using the setup from \cref{sec:photoevaporation_test}.

\subsection{Dust entrainment properties}
\label{sec:dust_entrainment_properties}

\subsubsection{Grain size}
\label{sec:grain_size}

Given the assumption that $\rho_\text{grain}$ is the same for all grain sizes, \cref{eq:drag_constant} shows that drag, and therefore dust entrainment, is inversely proportional to $s$. However, the transition from perfect entrainment in winds to perfect settling in discs is not well understood. To understand this better, we select five grain sizes, $s = [0.1,0.5,1,5,10]\,\mu$m, that bridge this transition. Because $0.1\,\mu$m grains are nearly perfectly coupled to the gas, we use their velocity and density to gauge the entrainment experienced by other grain sizes. \Cref{fig:grainsize} compares the dust densities for these grains after $\sim \! 10$ orbits at $5\,$AU. As expected, we see a steady decline in entrainment with increasing grain size, with a sharp change near $1\,\mu$m. By $5\,\mu$m, entrainment has ceased altogether. These results modify previous predictions of the maximum entrainable grain size in photoevaporative winds first made by \citet{Takeuchi/Clarke/Lin/2005} and later refined by \citet{Owen/Ercolano/Clarke/2011a}. Our results highlight the fact that the maximum grain size, at least for EUV photoevaporation, is limited by dust settling in the disc rather than aerodynamic drag in the wind. Unless mixing is forced, e.g. like in the \textsc{dustyphoto} test or the model used by \citet{Owen/Ercolano/Clarke/2011a}, larger grains will settle below the ionisation front such that they can no longer be dragged into the flow---even when the winds are physically able to support them. It is important to note that the $5\,\mu$m panel shows a small amount of dust in the outflow region. This dust results from numerical oscillations---the same post-shock oscillations we see in the zero drag \textsc{dustyshock} test (for a discussion on these oscillations, see \cref{sec:numerical_instability}). 

\subsubsection{Base flow density}
\label{sec:radiation flux}

We study the dust's sensitivity to the base flow density at the disc/wind interface because this is the only location in which photoevaporation can extract dust from the disc. Physically, the base flow density is determined by the density structure of the disc and the optical depth of ionising radiation. In our model, these properties are controlled by the midplane density, $\rho_\text{g,0}$, and the relative penetration depth, $\xi$. 

Our fiducial value of $\xi = 10^{-5}$ is a conservative estimate marking the lower boundary of where we expect EUV photoevaporation to take place (see \cref{fig:density_ts}). Variability in young stars can sustain EUV penetration depths at higher densities, although both FUV and X-ray radiation are significantly more penetrating. Our flexibility in setting $\xi$ allows us to explore these higher density regimes to approximate what dust entrainment is like for these deeper penetrating energies. Despite the very different heating mechanisms, once the gas is heated they all behave similarly hydrodynamically. Lower temperatures ($100$'s--$1000$'s K), non-isothermal flow, and largely neutral particles ($\mu \sim 2.3$) result in a lower sound speed, lower initial outflow velocity, and higher initial wind density. These are all multiplicative factors in the Epstein drag regime, and their combined effects (i.e. deviations from our `EUV' induced flow) will largely cancel out. Thus, aside from variations in heating caused by dust settling, we expect dust entrainment properties for FUV and X-ray induced flows to exhibit the same trends as the EUV case. As we are only interested in determining general trends at this point, we retain our EUV thermal switch as we explore the following (somewhat extreme) four penetration depths, $\xi = [10^{-4},10^{-3},10^{-2},10^{-1}]$, with grain sizes ranging from $0.1\,\mu$m to $1\,$m. 

\begin{figure}
	\centering{\includegraphics[width=\columnwidth]{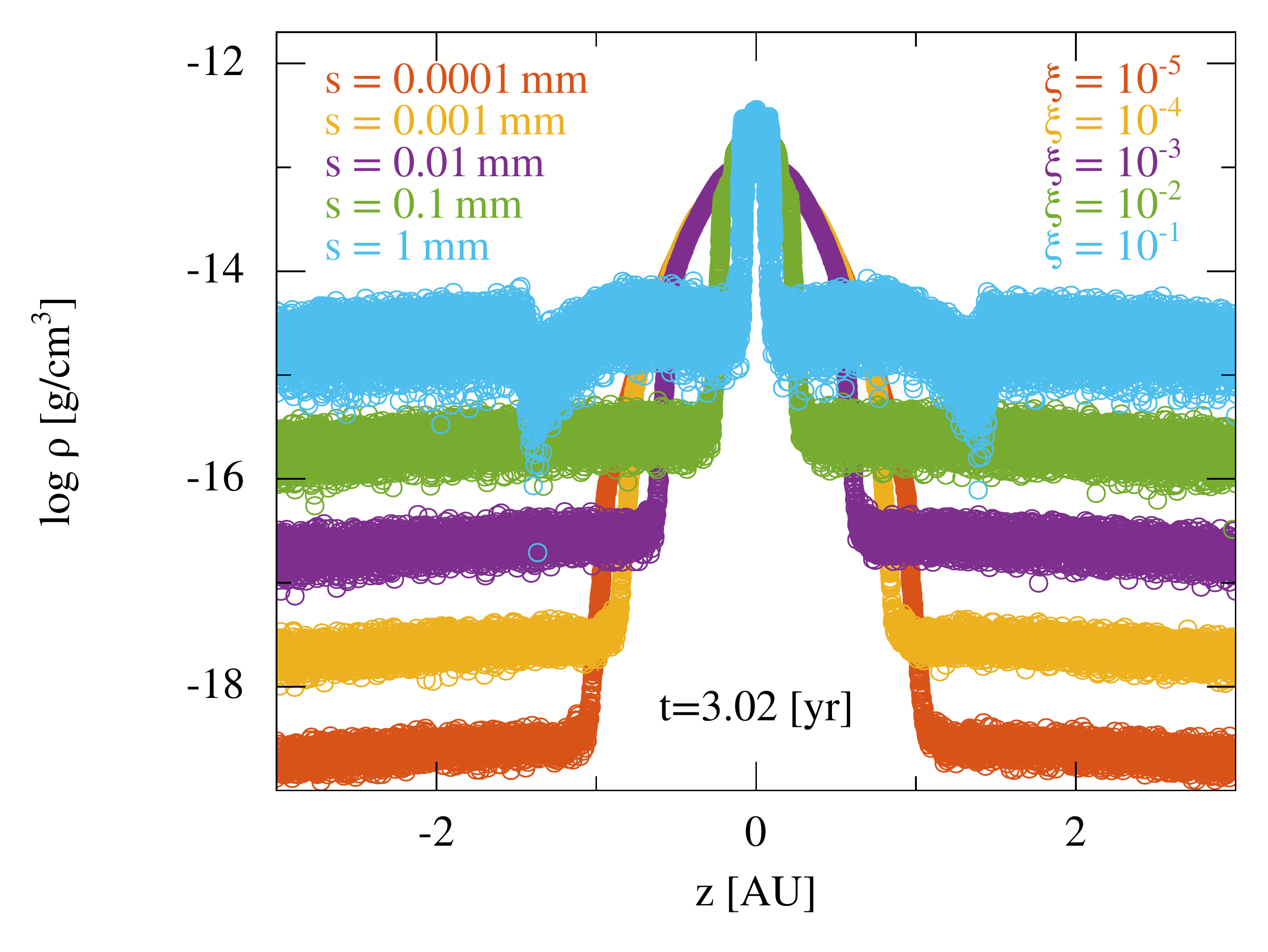}}
	\caption{Dust density of the largest well entrained grain size for each of five different penetration depths as a function of $z$. There is a linear correlation between the penetration depth $\xi$ and the maximally entrained grain size.}
	\label{fig:mindens_plot}
\end{figure}

For each different penetration depth, we observe the same general dependence on grain size we saw above, but the maximally entrained grain size increases linearly with $\xi$. This linear relationship can be seen in \cref{fig:mindens_plot}, which shows the density profile for the largest well-entrained grain size at each value of $\xi$. Importantly, \citet{Facchini/Clarke/Bisbas/2016} also find a linear relationship between the mass loss rate and the maximum grain size entrained by (external) FUV induced flows from the outer edges of discs. The similarity between our flows---despite having different heating mechanisms---supports our earlier claim that photoevaporative flows are hydrodynamically similar.

That we fail to see plumes of $\lesssim$mm grains emerging from protoplanetary discs suggests that extreme dispersal mechanisms, such as X-ray induced thermal sweeping \citep{Owen/Clarke/Ercolano/2012,Owen/etal/2013}, do not occur within $R\lesssim100\,$AU of the disc where these grains may still be present and midplane densities are large enough to entrain them. This supports the recent study by \citet{Haworth/Clarke/Owen/2016} who find that thermal sweeping only occurs in the tenuous outer edges of discs, if at all. We caution that any extreme dispersal mechanism will likely disrupt a large fraction of the dust in a way that is inconsistent with current observations.

The $\xi = [10^{-2},10^{-1}]$ simulations above are extreme photoevaporation scenarios and likely never to occur. Note the prominent dip in \cref{fig:mindens_plot} in the mm grains near $1.5\,$AU has nothing to do with entrainment, but rather shockwaves ringing through the disc as a result of the hydrostatic equilibrium of the disc being significantly disrupted. The combination of shock waves and high mass loss rates destroy the disc in a matter of years; hence the short timescale on our plot. In reality, we expect photoevaporation to be much more quiescent. A more physically motivated way of testing this density/grain size relationship would be to keep $\xi$ constant while increasing $\rho_\text{g,0}$. Not only does this maintain the vertical equilibrium of the disc, but it better reflects the fact that radiation flux and disc mass are loosely coupled by stellar mass---i.e. massive stars tend to have higher luminosities and higher disc masses. Returning to our fiducial penetration depth, $\xi = 10^{-5}$, we test $\rho_\text{g,0} = [10^{-12},10^{-11},10^{-10},10^{-9}]\,$g/cm$^3$. Because we want to isolate the coupling between base flow density and dust entrainment, we do not alter the mass of the central star just yet (see \cref{sec:stellar_mass}).

When we compare the results we obtain from varying $\rho_\text{g,0}$ to those obtained from varying $\xi$ above, we observe no perceptible differences in dust entrainment. This is surprising because the disc equilibrium is significantly different in both tests. The fact that the maximum entrained grain size is so simply related to the base flow density is fortunate for modellers. To illustrate, radiative transfer models that calculate the base flow gas density as a matter of course could use this relation to approximate the dust content in single-phase photoevaporative winds without requiring the complex two-phase hydrodynamic simulations we do here.

\subsubsection{Stellar mass}
\label{sec:stellar_mass}

\begin{figure}
	\centering{\includegraphics[width=\columnwidth]{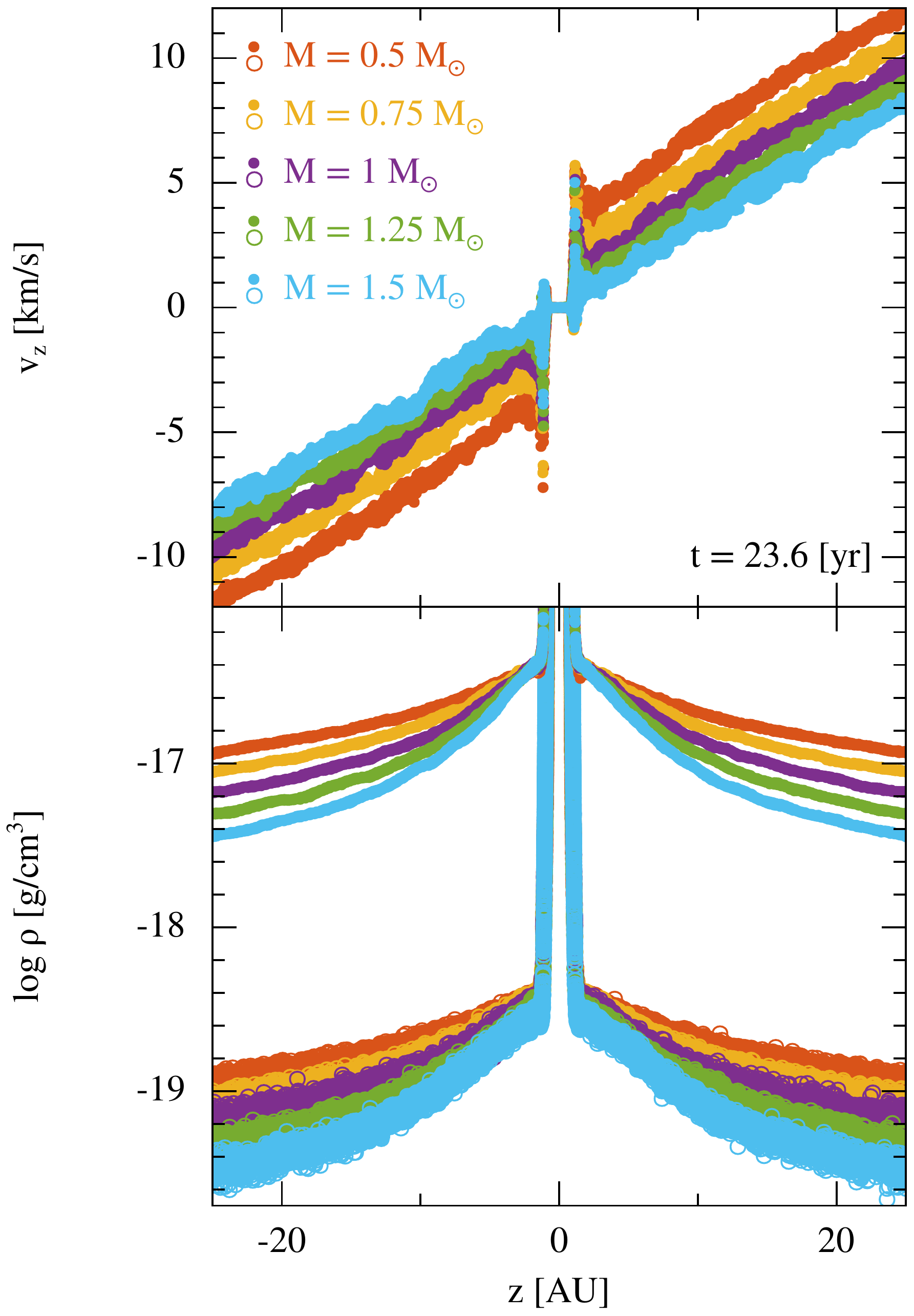}}
	\caption{Velocity (top) and density (bottom) profiles for gas (solid dots) and $0.1\,\mu$m dust grains (open circles) at five different stellar masses. The gas velocity is reduces as stellar mass increases, thus reducing the efficiency of dust entrainment. }
	\label{fig:mass_plot}
\end{figure}

As mentioned above, the stellar mass is not independent of disc mass or radiation luminosity. In the treatment above, we isolate and vary $\rho_\text{g,0}$ and $\xi$ individually while the stellar mass is fixed. We now reverse their roles and vary stellar mass while $\rho_\text{g,0}=10^{-11}\,\text{g/cm}^3$ and $\xi=10^{-5}$ remain fixed. Using five different stellar masses, $M=[0.5,0.75,1,1.25,1.5]\,M_\odot$, \cref{fig:mass_plot} shows the resulting velocity (top panel) and density (bottom panel) profiles for both gas and $s=0.1\,\mu$m dust grains. This time the gas velocity profile is altered in addition to its density because the escape speed is proportional to $\sqrt{M}$, 
\begin{equation}
	v_\text{esc} \approx \sqrt{\frac{2\mathcal{G}M}{R}}.
\end{equation}
This square-root dependence on mass can be observed in the relative spacing between wind velocities in the top panel.

Lower outflow velocities translate into steeper gradients in the gas density. The combination of low outflow velocities and low wind densities leads to a significant reduction in dust entrainment as $M$ increases. Even our ``perfectly''  entrained $0.1\,\mu$m grains show signs of waning entrainment by the way the base flow density for the dust is shifted successively lower with increasing $M$. Ultimately, this means that stellar mass competes with the synergistic effects of disc mass and luminosity to determine the entrainment properties of the flow.

\subsubsection{Distance to the central star}
\label{sec:radius}
\begin{figure*}
	\centering{\includegraphics[width=\textwidth]{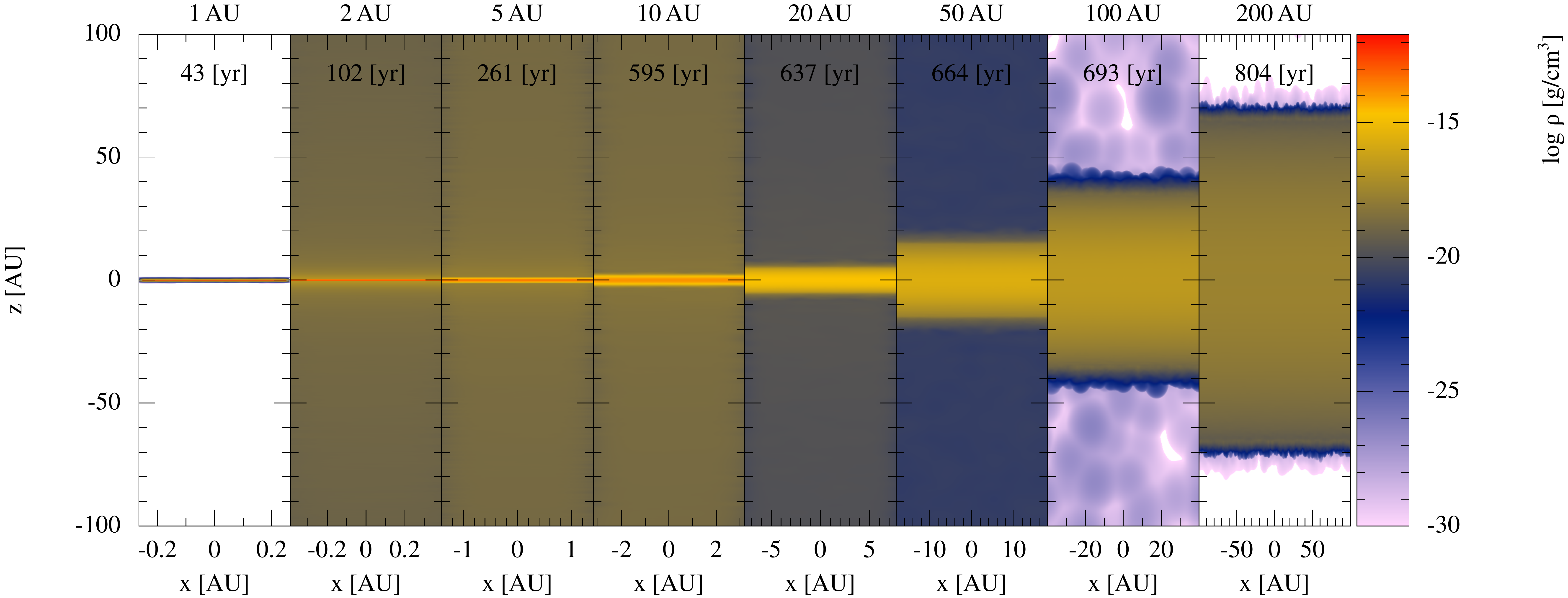}}
	\caption{Dust densities for $0.1\,\mu$m grains at eight radii in the disc spanning $1$--$200\,$AU. Photoevaporation and dust entrainment are most effective in the inner disc, but cannot operate at $R \lesssim 2\,$AU due to the strength of gravity. At large radii, photoevaporation continues to function, but the gas densities are too low to entrain even very small grains. We find there are three maxima of interest in the entrained dust phase: (i) the global density peak that occurs near the disc surface at the inner cutoff radius, (ii) the peak mass loss that occurs near $r_\text{g} \approx 9\,$AU, and (iii) the peak grain size entrained in the flow that occurs between $r_\text{g}$ and $2\,r_\text{g}$.}
	\label{fig:physical_radius}
\end{figure*}

We use our thin disc model to analyse how dust entrainment varies as a function of distance to the central star by obtaining the complete 2D density and thermal structure for our disc from an external model. Using the standard model from \citet{Woitke/etal/2016}, we select disc parameters from eight radii and insert these into our model. Because our thermal energy switch is comparatively simpler, we approximate the location of the ionisation front from their model. We also keep our ionisation temperature fixed at $10^4\,$K regardless of radius. The initial disc parameters we use for each radius are listed in \cref{tab:radial_init_cond}.
\begin{table}
	\centering
	\caption{Initial disc parameters at select radii are taken from the standard model of \citet{Woitke/etal/2016} in order to ensure we have a consistent disc structure from one simulation to the next. Relevant quantities at each radius are (from left to right) scale height $H_0$, midplane gas density $\rho_\text{g,0}$, ionisation front location $z_\text{i}$, and penetration depth $\xi$.}
	\label{tab:radial_init_cond}
	\sisetup{output-decimal-marker = {.}}
	\begin{tabular*}{\columnwidth}{@{\extracolsep{\stretch{1}}} S[table-format=3.0] S[table-format=2.2] c S[table-format=2.1] c @{}} \toprule
		{$R\,$(AU)} 	&	{$H_0\,$(AU)}		&	$\rho_\text{g,0}\,$(g/cm$^3$)		&	{$z_\text{i}\,$(AU)}	&	$\xi$			 		\\\midrule
		1			&	0.05				&	$2.0 \times 10^{-10}$			&	0.2					&	$5 \times 10^{-4}$		\\
		2			&	0.11				&	$4.0 \times 10^{-11}$			&	0.5					&	$1 \times 10^{-4}$		\\
		5			&	0.32				&	$4.0 \times 10^{-12}$			&	1.4					&	$1 \times 10^{-4}$		\\
		10			&	0.71				&	$1.5 \times 10^{-12}$			&	3.0					&	$1 \times 10^{-4}$		\\
		20			&	1.60				&	$2.5 \times 10^{-13}$			&	7.2					&	$5 \times 10^{-4}$		\\
		50			&	4.70				&	$2.5 \times 10^{-14}$			&	21.0				&	$5 \times 10^{-4}$		\\
		100			&	11.00				&	$3.0 \times 10^{-15}$			&	46.0				&	$5 \times 10^{-4}$		\\
		200			&	25.00				&	$2.5 \times 10^{-16}$			&	98.0				&	$1 \times 10^{-4}$		\\\bottomrule
	\end{tabular*}
\end{table}

\Cref{fig:physical_radius} compares the $0.1\,\mu$m dust density as a function of radius. Immediately obvious is the complete absence of dust grains in the outflow region at $R\lesssim2\,$AU and $R\gtrsim100\,$AU. The inner cutoff is due to photoevaporation's local inability to provide enough energy for gas to reach escape velocities. At large radii, dusty outflows are quenched by the diminished base flow densities brought about by the radial attenuation in gas density. More subtly, comparing these results to simulations of larger grain sizes shows that these inner/outer critical radii are different for each grain size. As $s$ increases, the dust entrainment region shrinks to a point located between $10$--$20\,$AU, or in terms of the gravitational radius ($r_\text{g} \equiv \mathcal{G}M_\star/c_\text{s}^2 \approx 9\,$AU), between $r_\text{g}$ and $2\,r_\text{g}$. This is the location where the largest possible grain size can be entrained in the flow. This should not be confused with the peak dust density in the flow, which always occurs close to the disc surface at the inner cutoff radius. Careful study of \cref{fig:physical_radius} confirms that the peak dust density in the wind occurs near the surface of the disc at $R=2\,$AU. Close inspection also reveals that the subsequent dust density is lower than the relatively constant neighbouring densities at $R=5$--$10\,$AU by a factor of $\sim \! 1.5$. This trend of initially high densities having a steeper gradient in the wind is observed in all of our dust grains. 

\begin{figure}
	\centering{\includegraphics[width=\columnwidth]{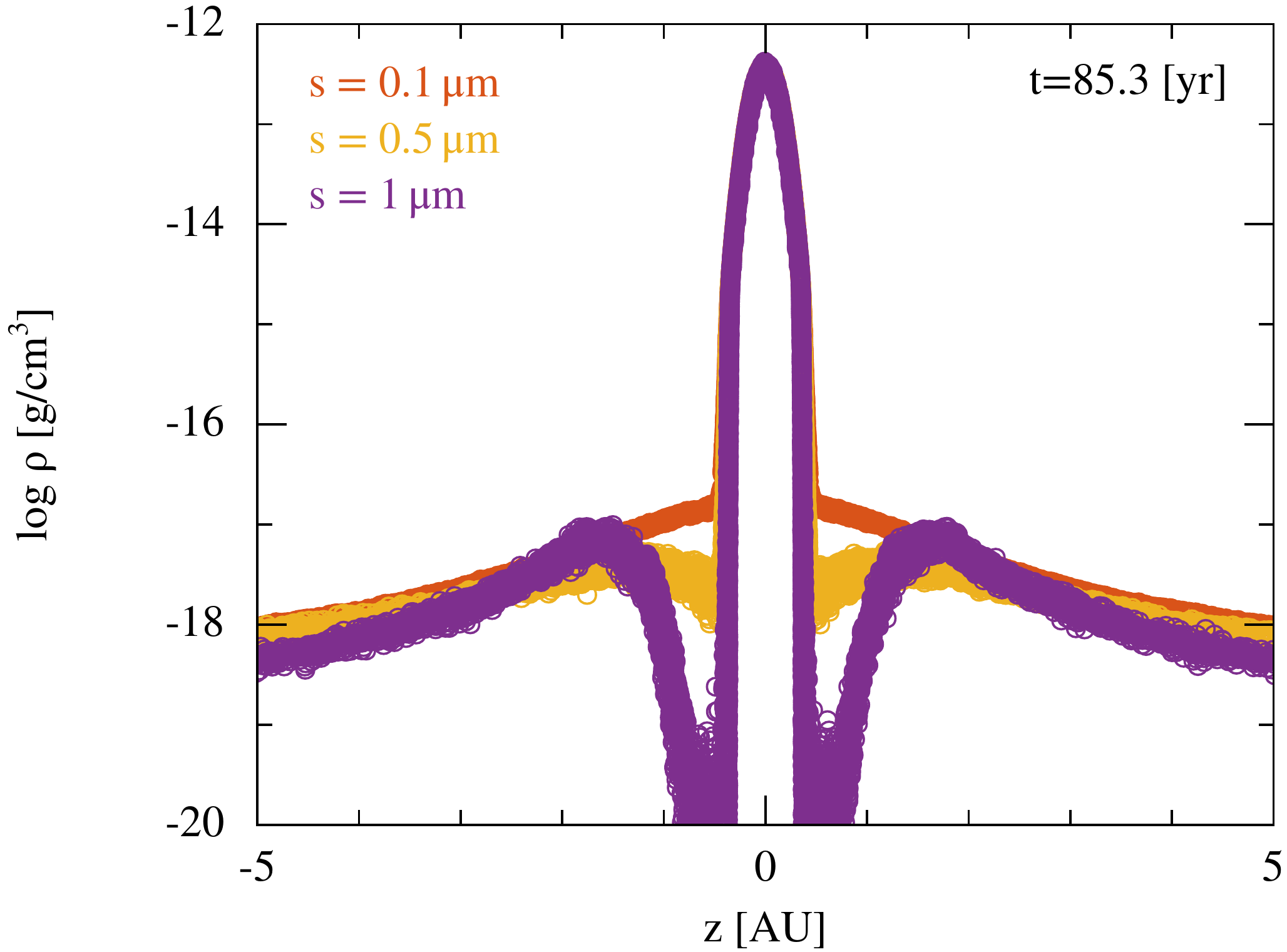}}
	\caption{Dust densities for three grain sizes are plotted as a function of z at $R=2\,$AU. Compared to the well entrained $0.1\,\mu$m grains, the weakly entrained $0.5\,\mu$m and $1\,\mu$m grains show progressive decoupling between dust in the disc and outflow. The static density peak in the outflow is created by dust that is trapped by the balance of (inward) gravitational settling and (outward) aerodynamic drag. The density trough, on the other hand, is created by quenching of dusty outflows through settling.}
	\label{fig:dust_cloud}
\end{figure}

These findings are in agreement with \citet{Owen/Ercolano/Clarke/2011a} who perform a similar study looking at maximally entrained grain sizes as a function of distance from the central star. An important difference between our studies, however, is that we are not restricted to simulating maximally entrained grains. It is easy to imagine a case where the grains are only marginally entrained and become decoupled from the gas in the low density outflow---we observe just such a scenario for $0.5\,\mu$m and $1\,\mu$m grains at $R=2\,$AU. The competition between gravitational settling and aerodynamic drag can trap intermediate sized grains in a quasi-steady dust cloud that hovers above the surface of the disc. Meanwhile grains in the disc continue to settle and evacuate the wind's launch region, creating a marked depletion of dust in the region above and below the ionisation front. As can be seen in \cref{fig:dust_cloud}, the depth, and width of this depleted region will depend strongly on grain size and probably radial location.
We expect the same phenomenon to occur at outer cutoff radii, but we do not observe any more of these dips---likely due to our sparse sampling in radius. The fact that we observe levitating dust grains in our outflows is interesting, but it is important to remember that these static concentrations of dust only occur in narrow regions around the cutoff radius for each grain size. It is uncertain whether these dust traps would persist in simulations of full discs. Mathematically, however, this is exactly the behaviour we would expect to see at critical points where the forces are balanced \citep[cf. the hover particles from][]{Liffman/Toscano/2000} and it shows that our algorithm can properly resolve the intermediate drag regime in photoevaporative outflows. Furthermore, it shows there is a richness in the dust dynamics that is only captured by fully modelling the coupled two-phase fluid equations for gas and dust.

\subsection{Plane-parallel approximation}
\label{sec:plane-parallel_approximation}

The basis for assuming vertical outflows from discs dates back to the seminal work of \citet{Hollenbach/etal/1994}, where they assume constant velocity, vertical outflows to simplify their treatment of photoevaporation. In the strong wind case, they claim that the flow propagates vertically to a height $z \sim R$ before pressure gradients bend the flow radially outwards. Later work by \citet{Font/etal/2004} provides more context by showing that the height of this vertical flow region is sensitive to the radial dependence in the base flow density profile. The shallower the density/pressure gradient, the more vertical the flow becomes. In fact, the remarkably vertical outflow seen in \citet{Alexander/Clarke/Pringle/2006a} is likely a result of decreasing pressure gradients on both sides of the the pressure maximum at the inner disc rim. As a result, our model may best be applied to regions near pressure maxima and/or discs with shallow surface density profiles.

\begin{figure}
	\centering{\includegraphics[width=\columnwidth]{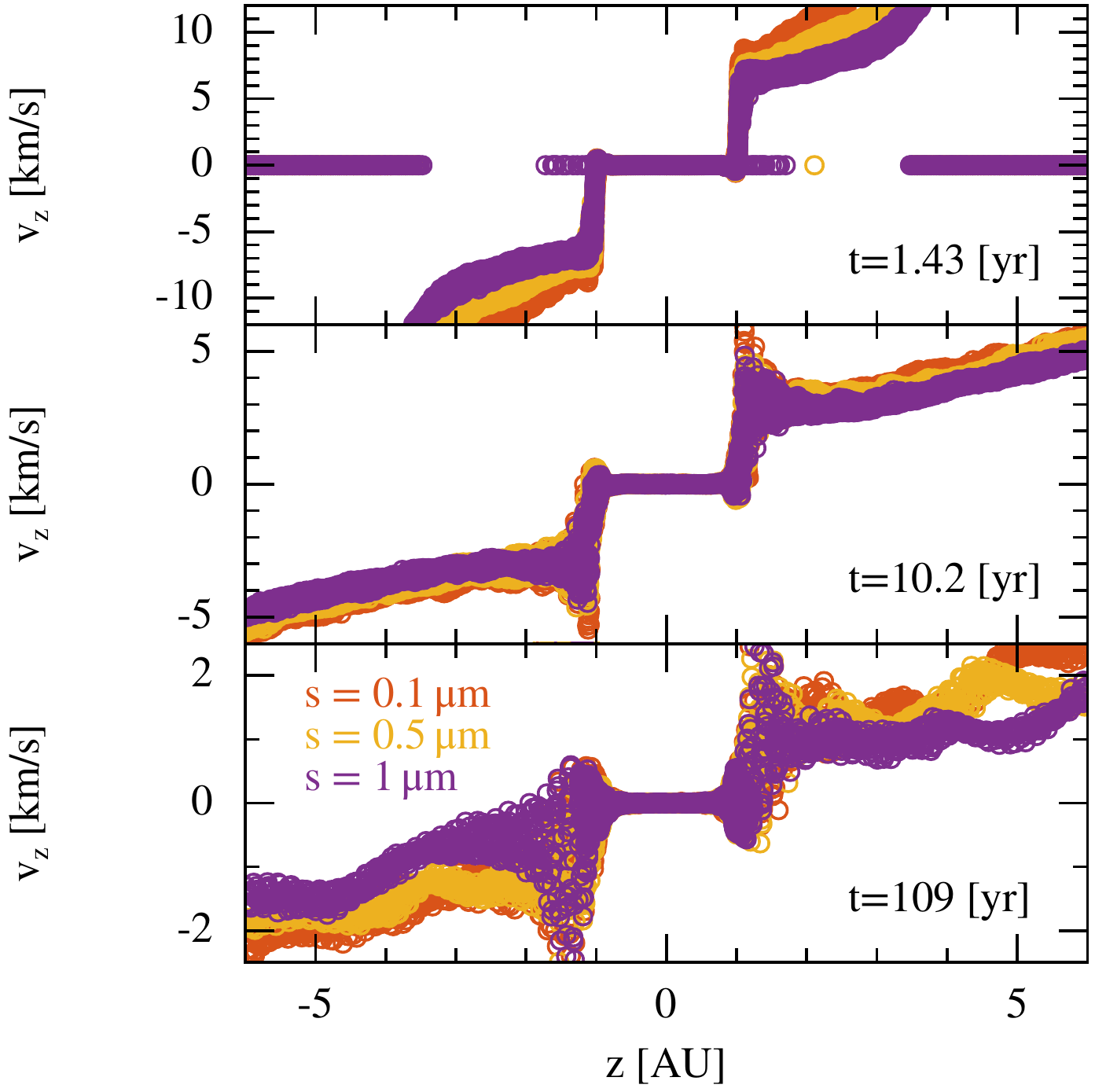}}
	\caption{Layering of dust velocities in phase space suggests that each grain size in \cref{fig:grainsize} will follow a unique trajectory when pulled into the wind. Initially, when the gas outflow is very fast, the different grain sizes experience significantly different outflow speeds. As time goes on and gas speeds drop, this layering of dust velocities---although still visible---becomes less distinct.}
	\label{fig:phase_grainsize}
\end{figure}

Lack of rotation is inherently assumed in vertical flow; however, the velocity profiles of dust grains in our model can provide some insight into the behaviour we expect to see for entrained dust grains in rotating outflows. Since larger dust grains feel less drag, we expect them to undergo less acceleration and, hence, have smaller velocities in the wind. Indeed, \cref{fig:phase_grainsize} shows that the three entrained grain sizes from \cref{sec:grain_size} exhibit a small, but definite layering in the dust velocities in the outflow. The layering is most pronounced initially when the gas velocities are high. Over time, the effect decreases as the outflow velocities settle into a steady state. In a vertical wind, grains with different outflow velocities are almost indistinguishable from one another, but in a rotating disc, the trajectories will separate due to conservation of angular momentum. Thus, any layering in phase space translates into stratification in altitude in position space, similar to projectile motion from a rotating object in a gravitational field. In extreme cases, the shallow trajectories from very slow grains could lead to recapture at large disc radii \citep{Clarke/Alexander/2016}\footnote{These authors also assume non-rotating, plane-parallel winds, but use a neat trick to add rotation back in for the dust. Our conjectures based on our 1D flow are consistent with their model.
}. Although we do not see a lot of evidence for such a transport mechanism due to such incremental changes in velocity and the sharp cutoff in entrained grain sizes in EUV flows due to settling, this effect should be further tested in a full 2D model with a proper thermal treatment of FUV and X-ray induced winds.

All of the observations made thus far depend in part on the velocity profile of the gas wind, which is underestimated in our model by assuming a non-rotating, plane-parallel geometry. Conservation of angular momentum causes steady-state photoevaporation streamlines to bend radially and diverge, the latter being the necessary condition allowing supersonic flow to occur in the wind \citep{Begelman/McKee/Shields/1983}. The lack of divergence in our vertical flow pushes the steady-state sonic point to $z=\pm \infty$, resulting in a velocity profile whose approximation to real photoevaporation declines with $z$ (our model is only designed to apply close to the disc surface). We overcome this initially by allowing our flow to expand into empty space, thereby allowing a sonic point to be established. However, over time the sonic point steadily moves outwards as the atmosphere above the disc fills with gas and back pressure retards the flow. The physical location of the sonic point is sensitive to the temperature profile of the wind, and hence the heating mechanism responsible for the flow. As direct absorption of Lyman continuum photons is the most efficient mechanism to heat the gas, EUV photoevaporation produces the closest sonic point to the disc (i.e. our largest underestimated velocity profile).

We can quantify how this underestimation affects dust entrainment by exploiting the observation above that a sonic point can be created in a vertical flow by manipulating the physical conditions above the disc. Note this artificial sonic point does not correct for the changing vertical gravitational field experienced during centrifugal expansion. However, because the streamlines are largely vertical and the contours of equal gravitational force move radially and vertically outward (for $z \leq \left. \pm R \middle/ \sqrt{2} \right.$), this effect is not crucially important in the region of interest. Sonic points ($z_\text{s}$) are established by introducing open boundary conditions in $z$. The open boundaries naturally set up a new, higher steady-state wind velocity by creating an infinite vacuum into which the fluid near the edge can continuously accelerate, but never fill up.

We trial two boundary positions at $z=\pm10$ and $\pm 3.5\,$AU (with sonic points located $\sim0.7\,$AU inside the boundary) corresponding to typical sonic surfaces found at $R=5\,$AU for X-ray and EUV photoevaporation, respectively \citep[see][]{Owen/Clarke/Ercolano/2012,Alexander/etal/2014}. In both cases, the increased velocity has no visible improvement on dust entrainment for the already perfectly-coupled $0.1\,\mu$m grains. In like manner, the $5$ and $10\,\mu$m grains remain just as oblivious to the wind as before. The only notable changes occur in the $0.5$ and $1\,\mu$m grains, whose dynamics and wind density can aptly be described by grains in our fiducial setup with proportionally smaller grain sizes. The proportionality factors for the artificial EUV and X-ray sonic points are, respectively, $\sim 0.5$ and $\sim 0.8$. Therefore, the downstream acceleration from conservation of angular momentum tends only to shift the dust entrainment properties to higher grain sizes with minimal effects outside the size range surrounding the maximum entrainable grain size. We may therefore conclude that the large and small dust grains in our model accurately determine the behaviour of dust in the disc and wind, while the intermediate sized grains can be used to provide a close lower bound on the entrainment properties of weakly-entrained grains near the ionisation front.

Finally, the new outflow velocity has little effect on the settling rate of grains in the disc. This suggests that turbulent mixing in the disc caused by the presence of the ionisation front is limited. As a result, the same grains that efficiently settle below the ionisation front in our fiducial model also settle here, yielding approximately the same maximum grain size in the wind. The difference is that the dust density for the maximum grain size is higher due to better coupling to the wind. This supports our claim that the maximum grain size entrained by EUV photoevaporation is set by dust settling in the disc rather than drag in the wind. Importantly, this result is independent of rotation and the geometry of the outflow. However, caution is required in extending this conclusion beyond EUV driven winds since the deeper penetration depths achieved by FUV and X-ray radiation could re-establish drag as the physical process limiting the grain size. We leave this for future studies to determine.

\subsection{Back-reaction on the gas}
\label{sec:backreaction_on_the_gas}

Given the small dust fraction in the wind, the back reaction on the gas is small, that is the velocity profile appears to be oblivious to the presence of the dust. We do, however, find a small decline in gas density as the entrained dust mass increases---approximately a $5\%$ decrease from $10\,\mu$m to $0.1\,\mu$m grains. This vertical offset in density has almost no $z$ dependence. Therefore, studies following the common practice of retroactively adding small, well-entrained dust grains to single-phase photoevaporative flows in order to do radiative transfer calculations should easily be able to account for this effect. Mass-loss calculations including dust should likewise be adjusted.

That we see a change in the gas density, but not velocity raises an interesting question: how much dust would it take to effect an observable change to the velocity profile? Understanding the effect of different dust-to-gas ratios is important because we have no definitive evidence that the canonical value of $0.01$ is physically correct in discs. In fact, photoevaporation is almost guaranteed to experience a range of dust-to-gas ratios as it clears gas from discs. To better understand how strongly photoevaporation may depend on the local dust fraction in the disc, we compare gas flow properties in discs with dust-to-gas ratios of $0.05,\text{ }0.1,\text{ and } 0.2$. Even at unreasonably high values like $0.2$, the reduction in outflow speed caused by dust entrainment is marginal, of order a few percent. At this point, however, the very structure of the gas disc is compromised by having to support so much additional mass against gravitational collapse. Therefore, we may safely assume that density is the only gas quantity significantly affected by the presence of dust in the flow.

A caveat to all of our simulations is that we do not couple our dust dynamics to the radiation from the central star. There are two main points to consider. The most important is that the opacity from entrained dust grains will reduce the radiation that would otherwise be captured by the disc, thereby altering the location of the ionisation front. As shown previously by varying $\rho_\text{i}$, this can have a profound effect upon both gas and dust properties in the wind. The other point to consider is that entrained dust grains will be exposed to radiation pressure from the central star. We do not expect radiation forces to be as important to photoevaporation as they are to Wolf-Rayet stars \citep[see][]{Pistinner/Eichler/1995}; nevertheless it could enhance dynamical feedback on the gas. Both of these points are beyond the scope of this study.


\section{Conclusions}
\label{sec:conclusions}

We show that by using unequal-mass, one-fluid SPH we can accurately simulate two-phase fluid dynamics in highly stratified atmospheres. We pioneer this method by simulating dusty photoevaporation on local scales in protoplanetary discs using a thin, non-rotating, plane-parallel atmosphere. We run a suite of simulations varying grain size, base flow density, stellar mass, distance to the central star, outflow velocity, and dust-to-gas ratio. We recover some important results from the literature and obtain new insights into the behaviour of dust grains in and around disc winds. Our main conclusions can be summarised as follows:

(i) Dust entrainment in typical photoevaporative winds is limited to micron sized grains and smaller. The largest grain size in EUV driven winds is a factor of a few smaller than the maximum entrainable grain size due to dust in the disc settling below the ionisation front. Further modelling is required to determine whether this is true for FUV and X-ray driven winds as well.

(ii) The velocities of entrained dust grains are weakly dependent on grain size. This could result in layering of different sized dust grains in the wind, but provides little evidence that this can be used as a transport mechanism whereby weakly entrained dust particles are recaptured in the outer regions of the disc.

(iii) There is approximately a one-to-one relation between the maximally entrained grain size and the base flow density of the gas. This has important observational ramifications for models that rely on rapid gas dispersal mechanisms in discs.

(iv) The maximum entrainable grain size is carried out at a critical radius between $r_\text{g}$ and $2\,r_\text{g}$. All smaller grains have an inner (outer) cutoff radius that is smaller (larger) than this critical radius. This means that photoevaporation cannot entrain dust grains in the very inner/outer regions of the disc.

(v) The peak dust density in the flow occurs close to the disc surface, near the inner cutoff radius. Flows with higher dust densities tend to have steeper gradients than those with lower dust densities.

(vi) In narrow regions surrounding each grain's cutoff radius, hovering dust grains concentrate and break away from similar sized grains in the disc that settle towards the midplane.

(vii) The back-reaction of the dust on the gas only has a measurable effect on the gas density. The gas velocity is only affected at extremely high dust-to-gas ratios.

Our non-rotating, plane-parallel photoevaporation model is ideal for developing intuition about dust dynamics in photoevaporative winds and obtaining local dust entrainment properties in their outflows. The numerical techniques we developed for this study could also be applied to global disc simulations, but would require addressing two additional numerical challenges. First, mixing of unequal-mass particles due to differential rotation and/or accretion, and second, implementing a radiative transfer mechanism suitable for SPH in multi-dimensions. Global two-phase simulations would not only further advance our understanding of the nuances of dusty photoevaporation, but could finally make photoevaporation models fully self-consistent with radiative transfer calculations by resolving the coupled feedback between disc heating, dust settling, and opacity from dusty winds. Furthermore, 3D simulations would allow us to explore photoevaporation in discs with non-axisymmetric features like spirals and warps induced by forming planets in the disc.

\section*{Acknowledgements} 
We would like to acknowledge Peter Woitke for meaningful discussions which resulted in an algorithm that was substantially more realistic and the anonymous referee for a prompt and useful report. Visualisations were performed using SPLASH \citep{Price/2007}. M.H. acknowledges funding from a Swinburne University Postgraduate Research Award (SUPRA). D.P. is supported by a Future Fellowship (FT130100034) from the Australian Research Council. G.L. acknowledges funding from the European Research Council for the FP7 ERC advanced grant project ECOGAL. This work was performed on the swinSTAR supercomputer at Swinburne University of Technology.




\bibliographystyle{mnras}
\bibliography{$HOME/Dropbox/Bibtex_library/library}

\begin{thebibliography}{}
\makeatletter
\relax
\def\mn@urlcharsother{\let\do\@makeother \do\$\do\&\do\#\do\^\do\_\do\%\do\~}
\def\mn@doi{\begingroup\mn@urlcharsother \@ifnextchar [ {\mn@doi@}
  {\mn@doi@[]}}
\def\mn@doi@[#1]#2{\def\@tempa{#1}\ifx\@tempa\@empty \href
  {http://dx.doi.org/#2} {doi:#2}\else \href {http://dx.doi.org/#2} {#1}\fi
  \endgroup}
\def\mn@eprint#1#2{\mn@eprint@#1:#2::\@nil}
\def\mn@eprint@arXiv#1{\href {http://arxiv.org/abs/#1} {{\tt arXiv:#1}}}
\def\mn@eprint@dblp#1{\href {http://dblp.uni-trier.de/rec/bibtex/#1.xml}
  {dblp:#1}}
\def\mn@eprint@#1:#2:#3:#4\@nil{\def\@tempa {#1}\def\@tempb {#2}\def\@tempc
  {#3}\ifx \@tempc \@empty \let \@tempc \@tempb \let \@tempb \@tempa \fi \ifx
  \@tempb \@empty \def\@tempb {arXiv}\fi \@ifundefined
  {mn@eprint@\@tempb}{\@tempb:\@tempc}{\expandafter \expandafter \csname
  mn@eprint@\@tempb\endcsname \expandafter{\@tempc}}}

\bibitem[\protect\citeauthoryear{{Adams}, {Hollenbach}, {Laughlin}  \&
  {Gorti}}{{Adams} et~al.}{2004}]{Adams/etal/2004}
{Adams} F.~C.,  {Hollenbach} D.,  {Laughlin} G.,   {Gorti} U.,  2004, \mn@doi
  [\apj] {10.1086/421989}, \href
  {http://adsabs.harvard.edu/abs/2004ApJ...611..360A} {611, 360}

\bibitem[\protect\citeauthoryear{{Alexander}, {Clarke}  \&
  {Pringle}}{{Alexander} et~al.}{2004}]{Alexander/Clarke/Pringle/2004b}
{Alexander} R.~D.,  {Clarke} C.~J.,   {Pringle} J.~E.,  2004, \mn@doi [\mnras]
  {10.1111/j.1365-2966.2004.08161.x}, \href
  {http://adsabs.harvard.edu/abs/2004MNRAS.354...71A} {354, 71}

\bibitem[\protect\citeauthoryear{{Alexander}, {Clarke}  \&
  {Pringle}}{{Alexander} et~al.}{2006}]{Alexander/Clarke/Pringle/2006a}
{Alexander} R.~D.,  {Clarke} C.~J.,   {Pringle} J.~E.,  2006, \mn@doi [\mnras]
  {10.1111/j.1365-2966.2006.10293.x}, \href
  {http://adsabs.harvard.edu/abs/2006MNRAS.369..216A} {369, 216}

\bibitem[\protect\citeauthoryear{{Alexander}, {Pascucci}, {Andrews}, {Armitage}
   \& {Cieza}}{{Alexander} et~al.}{2014}]{Alexander/etal/2014}
{Alexander} R.,  {Pascucci} I.,  {Andrews} S.,  {Armitage} P.,   {Cieza} L.,
  2014, \mn@doi [Protostars and Planets VI]
  {10.2458/azu_uapress_9780816531240-ch021}, \href
  {http://adsabs.harvard.edu/abs/2014prpl.conf..475A} {pp 475--496}

\bibitem[\protect\citeauthoryear{{Armitage}}{{Armitage}}{2011}]{Armitage/2011}
{Armitage} P.~J.,  2011, \mn@doi [\araa] {10.1146/annurev-astro-081710-102521},
  \href {http://adsabs.harvard.edu/abs/2011ARA%26A..49..195A} {49, 195}

\bibitem[\protect\citeauthoryear{{Begelman}, {McKee}  \& {Shields}}{{Begelman}
  et~al.}{1983}]{Begelman/McKee/Shields/1983}
{Begelman} M.~C.,  {McKee} C.~F.,   {Shields} G.~A.,  1983, \mn@doi [\apj]
  {10.1086/161178}, \href {http://adsabs.harvard.edu/abs/1983ApJ...271...70B}
  {271, 70}

\bibitem[\protect\citeauthoryear{{Clarke} \& {Alexander}}{{Clarke} \&
  {Alexander}}{2016}]{Clarke/Alexander/2016}
{Clarke} C.~J.,  {Alexander} R.~D.,  2016, submitted to MNRAS

\bibitem[\protect\citeauthoryear{Corless, Gonnet, Hare, Jeffrey  \&
  Knuth}{Corless et~al.}{1996}]{Corless/etal/1996}
Corless R.,  Gonnet G.,  Hare D.,  Jeffrey D.,   Knuth D.,  1996, \mn@doi
  [Advances in Computational Mathematics] {10.1007/BF02124750}, 5, 329

\bibitem[\protect\citeauthoryear{{Dehnen} \& {Aly}}{{Dehnen} \&
  {Aly}}{2012}]{Dehnen/Aly/2012}
{Dehnen} W.,  {Aly} H.,  2012, \mn@doi [\mnras]
  {10.1111/j.1365-2966.2012.21439.x}, \href
  {http://adsabs.harvard.edu/abs/2012MNRAS.425.1068D} {425, 1068}

\bibitem[\protect\citeauthoryear{{Epstein}}{{Epstein}}{1924}]{Epstein/1924}
{Epstein} P.~S.,  1924, \mn@doi [Physical Review] {10.1103/PhysRev.23.710},
  \href {http://adsabs.harvard.edu/abs/1924PhRv...23..710E} {23, 710}

\bibitem[\protect\citeauthoryear{{Ercolano}, {Clarke}  \& {Drake}}{{Ercolano}
  et~al.}{2009}]{Ercolano/Clarke/Drake/2009}
{Ercolano} B.,  {Clarke} C.~J.,   {Drake} J.~J.,  2009, \mn@doi [\apj]
  {10.1088/0004-637X/699/2/1639}, \href
  {http://adsabs.harvard.edu/abs/2009ApJ...699.1639E} {699, 1639}

\bibitem[\protect\citeauthoryear{{Facchini}, {Clarke}  \& {Bisbas}}{{Facchini}
  et~al.}{2016}]{Facchini/Clarke/Bisbas/2016}
{Facchini} S.,  {Clarke} C.~J.,   {Bisbas} T.~G.,  2016, \mn@doi [\mnras]
  {10.1093/mnras/stw240}, \href
  {http://adsabs.harvard.edu/abs/2016MNRAS.457.3593F} {457, 3593}

\bibitem[\protect\citeauthoryear{{Font}, {McCarthy}, {Johnstone}  \&
  {Ballantyne}}{{Font} et~al.}{2004}]{Font/etal/2004}
{Font} A.~S.,  {McCarthy} I.~G.,  {Johnstone} D.,   {Ballantyne} D.~R.,  2004,
  \mn@doi [\apj] {10.1086/383518}, \href
  {http://adsabs.harvard.edu/abs/2004ApJ...607..890F} {607, 890}

\bibitem[\protect\citeauthoryear{{Gingold} \& {Monaghan}}{{Gingold} \&
  {Monaghan}}{1977}]{Gingold/Monaghan/1977}
{Gingold} R.~A.,  {Monaghan} J.~J.,  1977, \mnras, \href
  {http://adsabs.harvard.edu/abs/1977MNRAS.181..375G} {181, 375}

\bibitem[\protect\citeauthoryear{{Glassgold}, {Najita}  \& {Igea}}{{Glassgold}
  et~al.}{1997}]{Glassgold/Najita/Igea/1997a}
{Glassgold} A.~E.,  {Najita} J.,   {Igea} J.,  1997, \apj, \href
  {http://adsabs.harvard.edu/abs/1997ApJ...480..344G} {480, 344}

\bibitem[\protect\citeauthoryear{{Gorti} \& {Hollenbach}}{{Gorti} \&
  {Hollenbach}}{2009}]{Gorti/Hollenbach/2009}
{Gorti} U.,  {Hollenbach} D.,  2009, \mn@doi [\apj]
  {10.1088/0004-637X/690/2/1539}, \href
  {http://adsabs.harvard.edu/abs/2009ApJ...690.1539G} {690, 1539}

\bibitem[\protect\citeauthoryear{{Haworth}, {Clarke}  \& {Owen}}{{Haworth}
  et~al.}{2016}]{Haworth/Clarke/Owen/2016}
{Haworth} T.~J.,  {Clarke} C.~J.,   {Owen} J.~E.,  2016, \mn@doi [\mnras]
  {10.1093/mnras/stv3016}, \href
  {http://adsabs.harvard.edu/abs/2016MNRAS.457.1905H} {457, 1905}

\bibitem[\protect\citeauthoryear{{Hollenbach}, {Johnstone}, {Lizano}  \&
  {Shu}}{{Hollenbach} et~al.}{1994}]{Hollenbach/etal/1994}
{Hollenbach} D.,  {Johnstone} D.,  {Lizano} S.,   {Shu} F.,  1994, \mn@doi
  [\apj] {10.1086/174276}, \href
  {http://adsabs.harvard.edu/abs/1994ApJ...428..654H} {428, 654}

\bibitem[\protect\citeauthoryear{{Hu} \& {Adams}}{{Hu} \&
  {Adams}}{2006}]{Hu/Adams/2006}
{Hu} X.~Y.,  {Adams} N.~A.,  2006, \mn@doi [Journal of Computational Physics]
  {10.1016/j.jcp.2005.09.001}, \href
  {http://adsabs.harvard.edu/abs/2006JCoPh.213..844H} {213, 844}

\bibitem[\protect\citeauthoryear{{Hutchison} \& {Laibe}}{{Hutchison} \&
  {Laibe}}{2016}]{Hutchison/Laibe/2016}
{Hutchison} M.~A.,  {Laibe} G.,  2016, \mn@doi [\pasa] {10.1017/pasa.2016.10},
  \href {http://adsabs.harvard.edu/abs/2016PASA...33...14H} {33, e014}

\bibitem[\protect\citeauthoryear{{Hutchison}, {Price}, {Laibe}  \&
  {Maddison}}{{Hutchison} et~al.}{2016}]{Hutchison/etal/2016b}
{Hutchison} M.~A.,  {Price} D.~J.,  {Laibe} G.,   {Maddison} S.~T.,  2016,
  submitted to MNRAS

\bibitem[\protect\citeauthoryear{{Koshizuka}, {Nobe}  \& {Oka}}{{Koshizuka}
  et~al.}{1998}]{Koshizuka/Nobe/Oka/1998}
{Koshizuka} S.,  {Nobe} A.,   {Oka} Y.,  1998, \mn@doi [International Journal
  for Numerical Methods in Fluids]
  {10.1002/(SICI)1097-0363(19980415)26:7<751::AID-FLD671>3.3.CO;2-3}, \href
  {http://adsabs.harvard.edu/abs/1998IJNMF..26..751K} {26, 751}

\bibitem[\protect\citeauthoryear{{Laibe} \& {Price}}{{Laibe} \&
  {Price}}{2011}]{Laibe/Price/2011}
{Laibe} G.,  {Price} D.~J.,  2011, \mn@doi [\mnras]
  {10.1111/j.1365-2966.2011.19291.x}, \href
  {http://adsabs.harvard.edu/abs/2011MNRAS.418.1491L} {418, 1491}

\bibitem[\protect\citeauthoryear{{Laibe} \& {Price}}{{Laibe} \&
  {Price}}{2012a}]{Laibe/Price/2012a}
{Laibe} G.,  {Price} D.~J.,  2012a, \mn@doi [\mnras]
  {10.1111/j.1365-2966.2011.20202.x}, \href
  {http://adsabs.harvard.edu/abs/2012MNRAS.420.2345L} {420, 2345}

\bibitem[\protect\citeauthoryear{{Laibe} \& {Price}}{{Laibe} \&
  {Price}}{2012b}]{Laibe/Price/2012b}
{Laibe} G.,  {Price} D.~J.,  2012b, \mn@doi [\mnras]
  {10.1111/j.1365-2966.2011.20201.x}, \href
  {http://adsabs.harvard.edu/abs/2012MNRAS.420.2365L} {420, 2365}

\bibitem[\protect\citeauthoryear{{Laibe} \& {Price}}{{Laibe} \&
  {Price}}{2014a}]{Laibe/Price/2014a}
{Laibe} G.,  {Price} D.~J.,  2014a, \mn@doi [\mnras] {10.1093/mnras/stu355},
  \href {http://adsabs.harvard.edu/abs/2014MNRAS.440.2136L} {440, 2136}

\bibitem[\protect\citeauthoryear{{Laibe} \& {Price}}{{Laibe} \&
  {Price}}{2014b}]{Laibe/Price/2014b}
{Laibe} G.,  {Price} D.~J.,  2014b, \mn@doi [\mnras] {10.1093/mnras/stu359},
  \href {http://adsabs.harvard.edu/abs/2014MNRAS.440.2147L} {440, 2147}

\bibitem[\protect\citeauthoryear{{Laibe} \& {Price}}{{Laibe} \&
  {Price}}{2014c}]{Laibe/Price/2014c}
{Laibe} G.,  {Price} D.~J.,  2014c, \mn@doi [\mnras] {10.1093/mnras/stu1367},
  \href {http://adsabs.harvard.edu/abs/2014MNRAS.444.1940L} {444, 1940}

\bibitem[\protect\citeauthoryear{{Laibe}, {Gonzalez}  \& {Maddison}}{{Laibe}
  et~al.}{2012}]{Laibe/Gonzalez/Maddison/2012}
{Laibe} G.,  {Gonzalez} J.-F.,   {Maddison} S.~T.,  2012, \mn@doi [\aap]
  {10.1051/0004-6361/201015349}, \href
  {http://adsabs.harvard.edu/abs/2012A%26A...537A..61L} {537, A61}

\bibitem[\protect\citeauthoryear{{Liffman} \& {Toscano}}{{Liffman} \&
  {Toscano}}{2000}]{Liffman/Toscano/2000}
{Liffman} K.,  {Toscano} M.,  2000, in Lunar and Planetary Institute Science
  Conference Abstracts. p.~1108

\bibitem[\protect\citeauthoryear{{Lodato} \& {Price}}{{Lodato} \&
  {Price}}{2010}]{Lodato/Price/2010}
{Lodato} G.,  {Price} D.~J.,  2010, \mn@doi [\mnras]
  {10.1111/j.1365-2966.2010.16526.x}, \href
  {http://adsabs.harvard.edu/abs/2010MNRAS.405.1212L} {405, 1212}

\bibitem[\protect\citeauthoryear{{Lor{\'e}n-Aguilar} \&
  {Bate}}{{Lor{\'e}n-Aguilar} \& {Bate}}{2014}]{Loren-Aguilar/Bate/2014}
{Lor{\'e}n-Aguilar} P.,  {Bate} M.~R.,  2014, \mn@doi [\mnras]
  {10.1093/mnras/stu1173}, \href
  {http://adsabs.harvard.edu/abs/2014MNRAS.443..927L} {443, 927}

\bibitem[\protect\citeauthoryear{{Lucy}}{{Lucy}}{1977}]{Lucy/1977}
{Lucy} L.~B.,  1977, \mn@doi [\aj] {10.1086/112164}, \href
  {http://adsabs.harvard.edu/abs/1977AJ.....82.1013L} {82, 1013}

\bibitem[\protect\citeauthoryear{{Merlin}, {Buonomo}, {Grassi}, {Piovan}  \&
  {Chiosi}}{{Merlin} et~al.}{2010}]{Merlin/etal/2010}
{Merlin} E.,  {Buonomo} U.,  {Grassi} T.,  {Piovan} L.,   {Chiosi} C.,  2010,
  \mn@doi [\aap] {10.1051/0004-6361/200913514}, \href
  {http://adsabs.harvard.edu/abs/2010A%26A...513A..36M} {513, A36}

\bibitem[\protect\citeauthoryear{{Miura} \& {Glass}}{{Miura} \&
  {Glass}}{1982}]{Miura/Glass/1982}
{Miura} H.,  {Glass} I.~I.,  1982, \mn@doi [Royal Society of London Proceedings
  Series A] {10.1098/rspa.1982.0107}, \href
  {http://adsabs.harvard.edu/abs/1982RSPSA.382..373M} {382, 373}

\bibitem[\protect\citeauthoryear{{Monaghan}}{{Monaghan}}{2002}]{Monaghan/2002}
{Monaghan} J.~J.,  2002, \mn@doi [\mnras] {10.1046/j.1365-8711.2002.05678.x},
  \href {http://adsabs.harvard.edu/abs/2002MNRAS.335..843M} {335, 843}

\bibitem[\protect\citeauthoryear{{Monaghan} \& {Price}}{{Monaghan} \&
  {Price}}{2006}]{Monaghan/Price/2006}
{Monaghan} J.~J.,  {Price} D.~J.,  2006, \mn@doi [\mnras]
  {10.1111/j.1365-2966.2005.09783.x}, \href
  {http://adsabs.harvard.edu/abs/2006MNRAS.365..991M} {365, 991}

\bibitem[\protect\citeauthoryear{{Osterbrock} \& {Ferland}}{{Osterbrock} \&
  {Ferland}}{2006}]{Osterbrock/Ferland/2006}
{Osterbrock} D.~E.,  {Ferland} G.~J.,  2006, {Astrophysics of gaseous nebulae
  and active galactic nuclei}.
University Science Books

\bibitem[\protect\citeauthoryear{{Owen}, {Ercolano}, {Clarke}  \&
  {Alexander}}{{Owen} et~al.}{2010}]{Owen/etal/2010}
{Owen} J.~E.,  {Ercolano} B.,  {Clarke} C.~J.,   {Alexander} R.~D.,  2010,
  \mn@doi [\mnras] {10.1111/j.1365-2966.2009.15771.x}, \href
  {http://adsabs.harvard.edu/abs/2010MNRAS.401.1415O} {401, 1415}

\bibitem[\protect\citeauthoryear{{Owen}, {Ercolano}  \& {Clarke}}{{Owen}
  et~al.}{2011}]{Owen/Ercolano/Clarke/2011a}
{Owen} J.~E.,  {Ercolano} B.,   {Clarke} C.~J.,  2011, \mn@doi [\mnras]
  {10.1111/j.1365-2966.2010.17750.x}, \href
  {http://adsabs.harvard.edu/abs/2011MNRAS.411.1104O} {411, 1104}

\bibitem[\protect\citeauthoryear{{Owen}, {Clarke}  \& {Ercolano}}{{Owen}
  et~al.}{2012}]{Owen/Clarke/Ercolano/2012}
{Owen} J.~E.,  {Clarke} C.~J.,   {Ercolano} B.,  2012, \mn@doi [\mnras]
  {10.1111/j.1365-2966.2011.20337.x}, \href
  {http://adsabs.harvard.edu/abs/2012MNRAS.422.1880O} {422, 1880}

\bibitem[\protect\citeauthoryear{{Owen}, {Hudoba de Badyn}, {Clarke}  \&
  {Robins}}{{Owen} et~al.}{2013}]{Owen/etal/2013}
{Owen} J.~E.,  {Hudoba de Badyn} M.,  {Clarke} C.~J.,   {Robins} L.,  2013,
  \mn@doi [\mnras] {10.1093/mnras/stt1663}, \href
  {http://adsabs.harvard.edu/abs/2013MNRAS.436.1430O} {436, 1430}

\bibitem[\protect\citeauthoryear{{Pistinner} \& {Eichler}}{{Pistinner} \&
  {Eichler}}{1995}]{Pistinner/Eichler/1995}
{Pistinner} S.,  {Eichler} D.,  1995, \mn@doi [\apj] {10.1086/176491}, \href
  {http://adsabs.harvard.edu/abs/1995ApJ...454..404P} {454, 404}

\bibitem[\protect\citeauthoryear{{Price}}{{Price}}{2004}]{Price/2004}
{Price} D.~J.,  2004, PhD thesis, Univ. Cambridge, Cambridge, UK

\bibitem[\protect\citeauthoryear{{Price}}{{Price}}{2007}]{Price/2007}
{Price} D.~J.,  2007, \mn@doi [\pasa] {10.1071/AS07022}, \href
  {http://adsabs.harvard.edu/abs/2007PASA...24..159P} {24, 159}

\bibitem[\protect\citeauthoryear{{Price}}{{Price}}{2012}]{Price/2012}
{Price} D.~J.,  2012, \mn@doi [Journal of Computational Physics]
  {10.1016/j.jcp.2010.12.011}, \href
  {http://adsabs.harvard.edu/abs/2012JCoPh.231..759P} {231, 759}

\bibitem[\protect\citeauthoryear{{Price} \& {Federrath}}{{Price} \&
  {Federrath}}{2010}]{Price/Federrath/2010}
{Price} D.~J.,  {Federrath} C.,  2010, \mn@doi [\mnras]
  {10.1111/j.1365-2966.2010.16810.x}, \href
  {http://adsabs.harvard.edu/abs/2010MNRAS.406.1659P} {406, 1659}

\bibitem[\protect\citeauthoryear{{Price} \& {Laibe}}{{Price} \&
  {Laibe}}{2015}]{Price/Laibe/2015}
{Price} D.~J.,  {Laibe} G.,  2015, \mn@doi [\mnras] {10.1093/mnras/stv996},
  \href {http://adsabs.harvard.edu/abs/2015MNRAS.451..813P} {451, 813}

\bibitem[\protect\citeauthoryear{{Price} \& {Monaghan}}{{Price} \&
  {Monaghan}}{2007}]{Price/Monaghan/2007}
{Price} D.~J.,  {Monaghan} J.~J.,  2007, \mn@doi [\mnras]
  {10.1111/j.1365-2966.2006.11241.x}, \href
  {http://adsabs.harvard.edu/abs/2007MNRAS.374.1347P} {374, 1347}

\bibitem[\protect\citeauthoryear{{Rasio} \& {Lombardi}}{{Rasio} \&
  {Lombardi}}{1999}]{Rasio/Lombardi/1999}
{Rasio} F.~A.,  {Lombardi} Jr. J.~C.,  1999, Journal of Computational and
  Applied Mathematics, \href
  {http://adsabs.harvard.edu/abs/1999JCoAM.109..213R} {109, 213}

\bibitem[\protect\citeauthoryear{{Sod}}{{Sod}}{1978}]{Sod/1978}
{Sod} G.~A.,  1978, \mn@doi [Journal of Computational Physics]
  {10.1016/0021-9991(78)90023-2}, \href
  {http://adsabs.harvard.edu/abs/1978JCoPh..27....1S} {27, 1}

\bibitem[\protect\citeauthoryear{{Takeuchi}, {Clarke}  \& {Lin}}{{Takeuchi}
  et~al.}{2005}]{Takeuchi/Clarke/Lin/2005}
{Takeuchi} T.,  {Clarke} C.~J.,   {Lin} D.~N.~C.,  2005, \mn@doi [\apj]
  {10.1086/430393}, \href {http://adsabs.harvard.edu/abs/2005ApJ...627..286T}
  {627, 286}

\bibitem[\protect\citeauthoryear{{Tielens} \& {Hollenbach}}{{Tielens} \&
  {Hollenbach}}{1985}]{Tielens/Hollenbach/1985a}
{Tielens} A.~G.~G.~M.,  {Hollenbach} D.,  1985, \mn@doi [\apj]
  {10.1086/163111}, \href {http://adsabs.harvard.edu/abs/1985ApJ...291..722T}
  {291, 722}

\bibitem[\protect\citeauthoryear{Wendland}{Wendland}{1995}]{Wendland/1995}
Wendland H.,  1995, \mn@doi [Advances in Computational Mathematics]
  {10.1007/BF02123482}, 4, 389

\bibitem[\protect\citeauthoryear{{Woitke} et~al.,}{{Woitke}
  et~al.}{2016}]{Woitke/etal/2016}
{Woitke} P.,  et~al., 2016, \mn@doi [\aap] {10.1051/0004-6361/201526538}, \href
  {http://adsabs.harvard.edu/abs/2016A%26A...586A.103W} {586, A103}

\makeatother
\end{thebibliography}



\appendix

\section{Numerical tests of dust-gas numerical method}
\label{sec:standard_tests}

Incorporating all of the numerical algorithms needed to simulate photoevaporation and small dust grains together in a single code involved significant changes to our standard SPH code: (i) the one-fluid formalism, (ii) the variable smoothing length and number density formulation for unequal masses, (iii) the Wendland kernels, and (iv) the implicit timestepping. We also parallelised the code using OpenMP to improve speed. In order to ensure the code is working properly, we have benchmarked it using some canonical gas/dust test problems: \textsc{dustybox} and \textsc{dustywave} \citep{Laibe/Price/2011}, \textsc{dustyshock} \citep{Sod/1978,Miura/Glass/1982,Laibe/Price/2012a}, and the settling test \citep[hereafter \textsc{dustydisc};][]{Price/Laibe/2015}. These tests ensure that our numerical implementation of the equations in \cref{sec:sph_formalism} are correct and show that we are able to accurately track the aerodynamic coupling between gas and dust in very high drag regimes. 

\subsection{\sc{dustybox}}
\label{sec:dustybox}

The \textsc{dustybox} problem is a simple drag test using a two-phase fluid of gas and dust at rest in their barycentric frame of reference, i.e. $\mathbf{v} = 0$, but where $\Delta \mathbf{v} \neq 0$. This test isolates and checks the drag terms since all other terms are zero. More importantly, it verifies our implicit time-stepping algorithm in \cref{sec:time-stepping} because it uses the same analytic solution. If done correctly, the solutions should match exactly.

We set $100$ equally spaced particles in a periodic 1D box, $ x \in \left[ -0.5 , 0.5 \right] $. The total density and sound speed of the gas are  $\rho = c_\text{s} =1$, the dust fraction is $\epsilon = 0.5$, and the dust velocity $\mathbf{v}_\text{d} = 0.01$, all in code units. No dissipation or viscosity is applied in this problem. We performed tests at five different drag regimes ($K = [0.01,0.1,1,10,100]$) and found that the implicit scheme reproduces the analytic velocity curves for the dust to machine precision. This shows that our proposed implicit evolution equations for $\Delta \mathbf{v}$ are valid. Furthermore, we confirmed that the energy for the system is exactly conserved (along with mass and momentum), implying that our evolution equations for $u$ are also sound.

\subsection{\sc{dustywave}}
\label{sec:dustywave}

\begin{figure}
\centering{\includegraphics[width=\columnwidth]{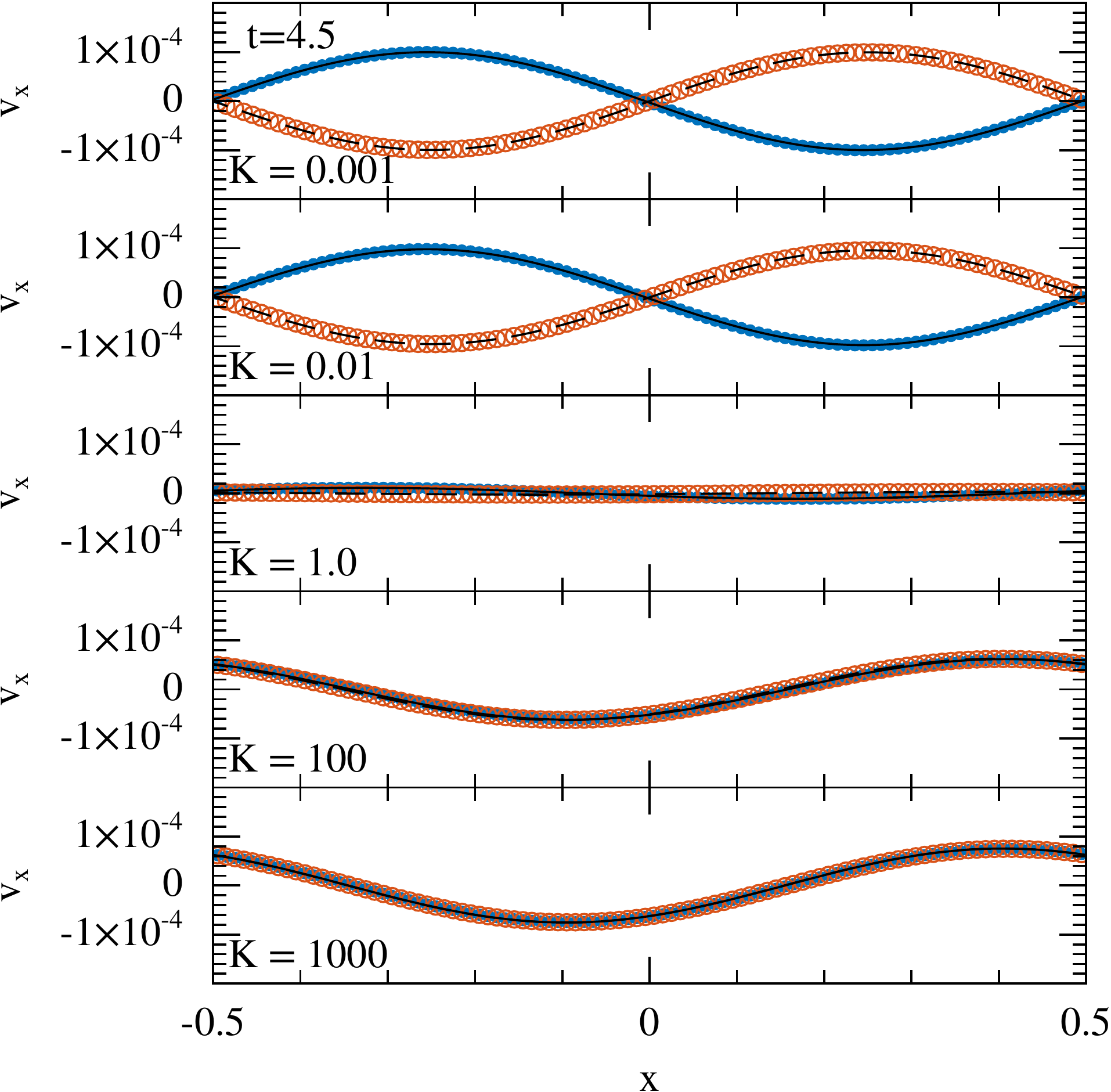}}
\caption{Gas (blue filled circles) and dust (brown open circles) velocities are plotted after $4.5$ periods in the \textsc{dustywave} problem using $100$ unequal-mass particles and the drag coefficient listed in each panel. The black curves are the exact solutions for the gas (solid) and dust (dashed). There are no adverse effects from using unequal masses in this test.}
\label{fig:wave_test_plot}
\end{figure}

The \textsc{dustywave} problem \citep{Laibe/Price/2011} probes the pressure force in the gas using the propagation of linear sound waves caused by small sinusoidal perturbations in the velocity, density, and energy. Instead of stretching equal-mass particles to produce the desired density perturbation, we simply build the perturbation directly into the particle masses and leave the particle spacing uniform. The equilibrium values describing the fluid are as follows: $v_0 = 0$, $\rho_\text{g,0} = 1$, and $u_0 = P_0/[(\gamma-1) \rho_\text{g,0}]$ with $\gamma = 5/3$ and $P_0 = 3/5$, such that $c_{\text{s},0} = 1$. The perturbation about each equilibrium, on the other hand, is uniform: $\delta_m = \delta_\rho = \delta_v = \delta_u = 10^{-4}$. Just as in \textsc{dustybox}, all quantities are in code units and there is no dissipation or viscosity applied to the system. Again we use $100$ particles, but this time we test a broader range in the drag coefficient ($K = [0.001,0.01,1, 100,1000]$), the results of which are shown in \cref{fig:wave_test_plot}. The $L_2$ errors as computed by \textsc{splash} \citep{Price/2007} are in every case less than $2\%$, consistent with a second-order integration scheme \citep[cf.][]{Laibe/Price/2014b}. These results are indistinguishable from those obtained using an equal-mass setup, implying that small mass differences are handled properly by our algorithm in \cref{sec:unequalmassparticles}

\subsection{\sc{dustyshock}}
\label{sec:dustyshock}

The \textsc{dustyshock} problem tracks a shock wave propagating through a dust/gas mixture initially created by discontinuities in the density and pressure. \textsc{dustyshock} is an important test, not because we expect to see strong shocks in our system, but because photoevaporation and/or dust settling can create steep gradients in the fluid variables, similar to a discontinuity in a shock front. Moreover, it is the first test problem in which we use the full set of SPH fluid equations, including viscosity, dissipation, conduction, and unequal masses. In other words, it tests our entire code except our thermal energy switch for photoevaporation.

\subsubsection{Setups}
\label{sec:dustyshock_setup}

The initial discontinuities are created by placing two fluids in a 1D box, $x \in \left[ -0.5, 0.5 \right]$, separated by a membrane at $x = 0$ (all quantities in code units). We use reflective boundary conditions to simulate fluids that extend to $\pm \infty$; this requires both fluids to start from rest and care must be taken to ensure the shock does not interact with the boundaries. The density and pressure of the gas on the ``left'' ($x<0$) and ``right'' ($x \geq 0$) are $\rho_\text{\tiny{L}} = P_\text{\tiny{L}} = 1$ and $\rho_\text{\tiny{R}} = P_\text{\tiny{R}} = 0.125$, respectively. Finally, \textsc{dustyshock} requires an adiabatic equation of state; we use $P = (\gamma - 1) \rho u$ with $\gamma = 5/3$. At $t = 0$, the membrane is removed and the fluids are allowed to move freely. The interacting fluids will pass through a non-linear transient phase, the duration of which will depend on the drag regime, and will eventually arrive at a stationary phase where the solution is the same as that of a pure gas fluid moving at a modified $\gamma$ and sound speed \citep{Miura/Glass/1982,Laibe/Price/2012a}. Unlike the previous tests, we are only interested in testing the very high and low (i.e. zero) drag regimes because there is no analytic solution for the transient phase. For the high drag regime we use a drag coefficient $K = 1000$ and a constant dust fraction $\epsilon = 0.5$. We test the zero-drag case using two different setups. The first setup is exactly the same as the high-drag case, while the second differs by having a constant dust density of $\rho_\text{d}=0.125$ across both fluids. This latter scenario requires a discontinuous jump in the dust fraction: $\epsilon_\text{\tiny{L}} = 1/9$ and $\epsilon_\text{\tiny{R}} = 0.5$.

\subsubsection{\sc{dustyshock} \textnormal{\emph{with unequal-mass fluids}}}
\label{sec:dustyshock_unequal_mass_fluids}

When using equal-mass particles, the density difference between the two fluids is created by using $8\times$ more particles on the left than on the right. Because we need a minimum of about $150$ particles in order to resolve the discontinuity of the forward ($x \geq 0$) propagating shock wave, we are required to use at least $1200$ particles for $x<0$---much greater than is required to sufficiently resolve the shock. This prohibitive condition can be relaxed completely if instead we create the density jump using unequal-mass particles. Then we can minimally resolve the entire shock using a total of only $300$ particles. Alternatively, running with $1350$ particles results in significantly improved resolution of the low-density fluid and even slightly faster execution times because the Courant condition is no longer being controlled by tightly spaced particles. However, a glitch in several fluid quantities occurs at the interface between the different mass particles, regardless of the drag regime. A similar feature---albeit much reduced---also appears in equal-mass tests, but is caused by the sudden change in particle spacing. Because our unequal-mass tests use equal/smoothly varying particle spacings, this may indicate that large mass differences between interacting particles can cause numerical irregularities.

To test our algorithm's sensitivity to jumps in the particle mass we decrease (increase) the density jump between fluids by progressively increasing (decreasing) the mass of the particles for $x \geq 0$. The width and amplitude of the aforementioned glitch steadily shrinks (grows) as the mass difference becomes less (more) pronounced. For example, a mass ratio of $2$ reduces the size of the glitch to that of the equal-mass case while mass ratios $\gtrsim120$ begin to exhibit severe departures from the analytic solution. By $128$ the solution becomes unstable. We found this is because the recoil of lower mass particles cannot properly be resolved within a Courant-limited timestep. Forcing a smaller timestep fixes the issue (e.g. one tenth the Courant timestep allows us to extend stability up to mass ratios of $\sim \!180$) which suggests there may be a mass-timestep criterion that must be met to ensure accuracy and stability in extreme cases where large mass differences between \emph{interacting} particles are required. As we do not anticipate such extreme interactions in our simulations, we leave this as a topic for future studies.

It is worth pointing out that the above \textsc{dustyshock} tests involve a very limited number of unequal-mass interactions. Although useful for probing our number density formulation, they are actually poor tests of how unequal-mass particles behave in simulations. Furthermore, because all of the unequal-mass encounters take place in the vicinity of the glitch at the contact discontinuity, there is a danger in misconstruing these tests as evidence that unequal-mass encounters produce universally worse results than their equal-mass counterparts. A better test of the behaviour of unequal-mass particles would be to give each particle a unique mass in a smoothly varying mass distribution.

\subsubsection{\sc dustyshock \textnormal{\emph{with unequal-mass particles}}}
\label{sec:dustyshock_mass_distribution}
\begin{figure*}
	\centering{\includegraphics[width=\columnwidth]{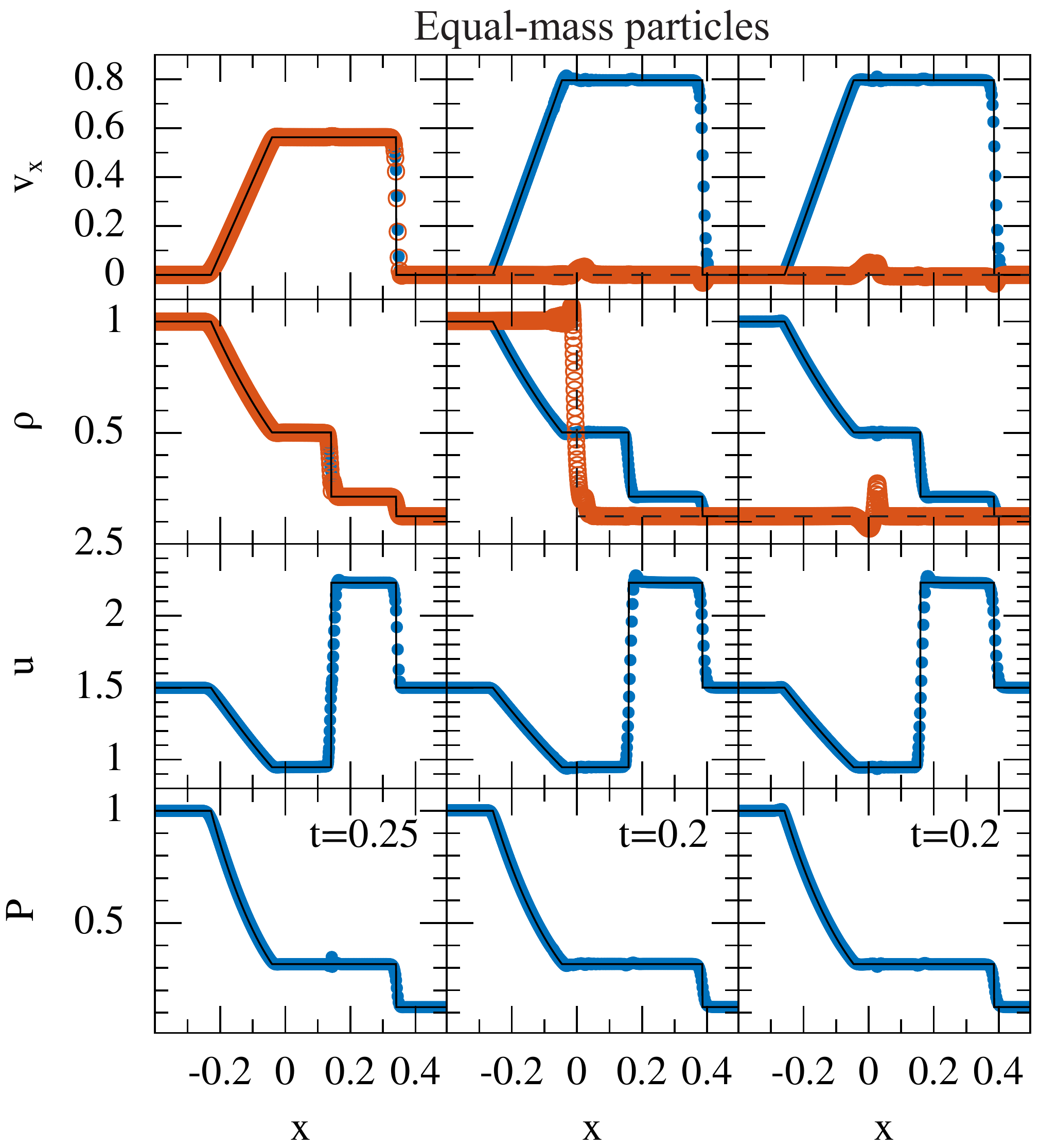}}
	\centering{\includegraphics[width=\columnwidth]{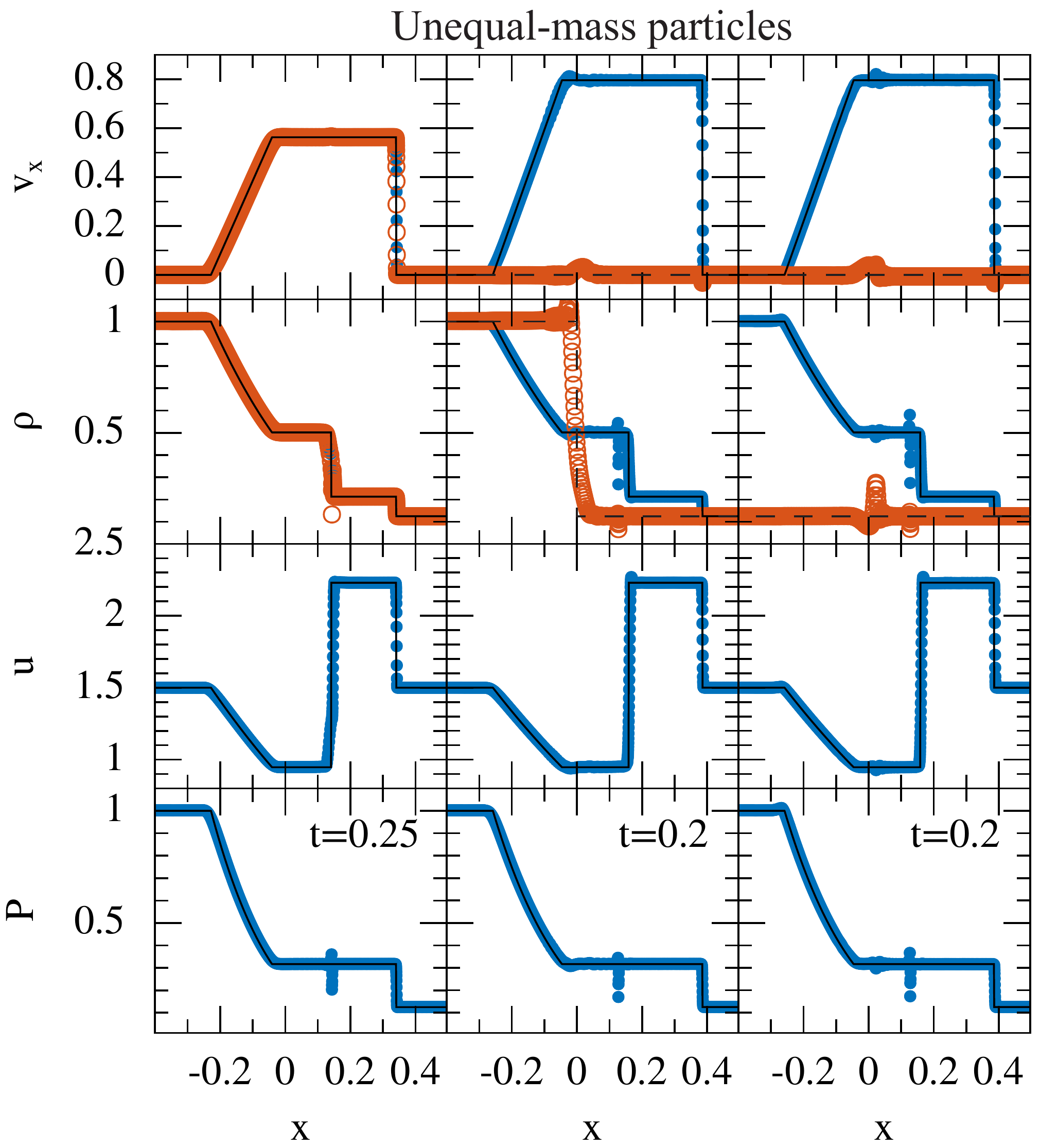}}
	\caption{Comparison of the \textsc{dustyshock} test with $1350$ particles using equal-mass particles (left) and unequal-mass particles (right). Rows (from top to bottom) show velocity, density, internal energy, and pressure. Columns separate different \textsc{dustyshock} setups (from left to right): $K=1000$, $K=0$ retaining the discontinuity in the dust density, and $K=0$ with a uniform dust density throughout. Gas quantities are represented by the solid (blue) dots, dust quantities use the open (brown) circles, and the black lines are the analytic solution for the gas (solid) and dust (dashed) in each of the respective plots. The tests run with unequal-mass particles have a notable irregularity at the contact discontinuity, but otherwise reproduce the solutions just as well or better than the equal-mass case due to increased resolution at the discontinuities.}
	\label{fig:shock_test_plot}
\end{figure*}

Because the fluid density is degenerately defined by both mass and particle spacing, there is nothing preventing us from building a constant density out of a smoothly varying mass distribution. In practice, this can be achieved by adapting the stretching technique often used to make the density perturbation in the \textsc{dustywave} problem \citep[see][]{Price/2004}. In this case, however, we cancel out the spatial perturbation by carefully assigning masses that keep the density constant. First, we note that the cumulative mass of the system is given by $M(x_a) = \rho_0 x_a$ and the individual masses by $m_a = \delta M(x_a) = \rho_0 \delta x_a$. Thus, the mass of each particle simply depends on the gradient of the number density. If we define $x_a \equiv f(x_{a,0})$, where $f$ is a stretching function of our choice acting on a set of uniformly-spaced particle coordinates $x_{a,0}$, then the mass of each particle is directly proportional to $f'(x_{a,0})$.

An additional complication arises because our reflective boundary conditions are inconsistent with a non-uniform mass distribution. We avoid this complication by converting a portion of the simulation particles near the boundaries of the computational domain into ghost particles, thereby redefining where our boundaries occur. Although this method can carve out a considerable portion of the computational domain at low resolution, the effect is minimal with $1350$ particles---the resolution we use to compare with the equal-mass case.

\Cref{fig:shock_test_plot} shows a side-by-side comparison of the three different \textsc{dustyshock} setups using equal masses (left) and a mass distribution with stretch function $f(x)=\sqrt{x}$ (right). Our only criteria for choosing $f$ was that it should exhibit a large range in \emph{relative} masses between particle neighbours. These tests show that unequal-mass particles behave just as well as equal-mass particles as long as the mass distribution is smooth and the relative mass difference between particle pairs is less than $\sim \!1$--$2$ orders of magnitude.

\subsection{\sc{dustydisc}}
\label{sec:dustydisc}
\begin{figure*}
	\centering{\includegraphics[width=\textwidth]{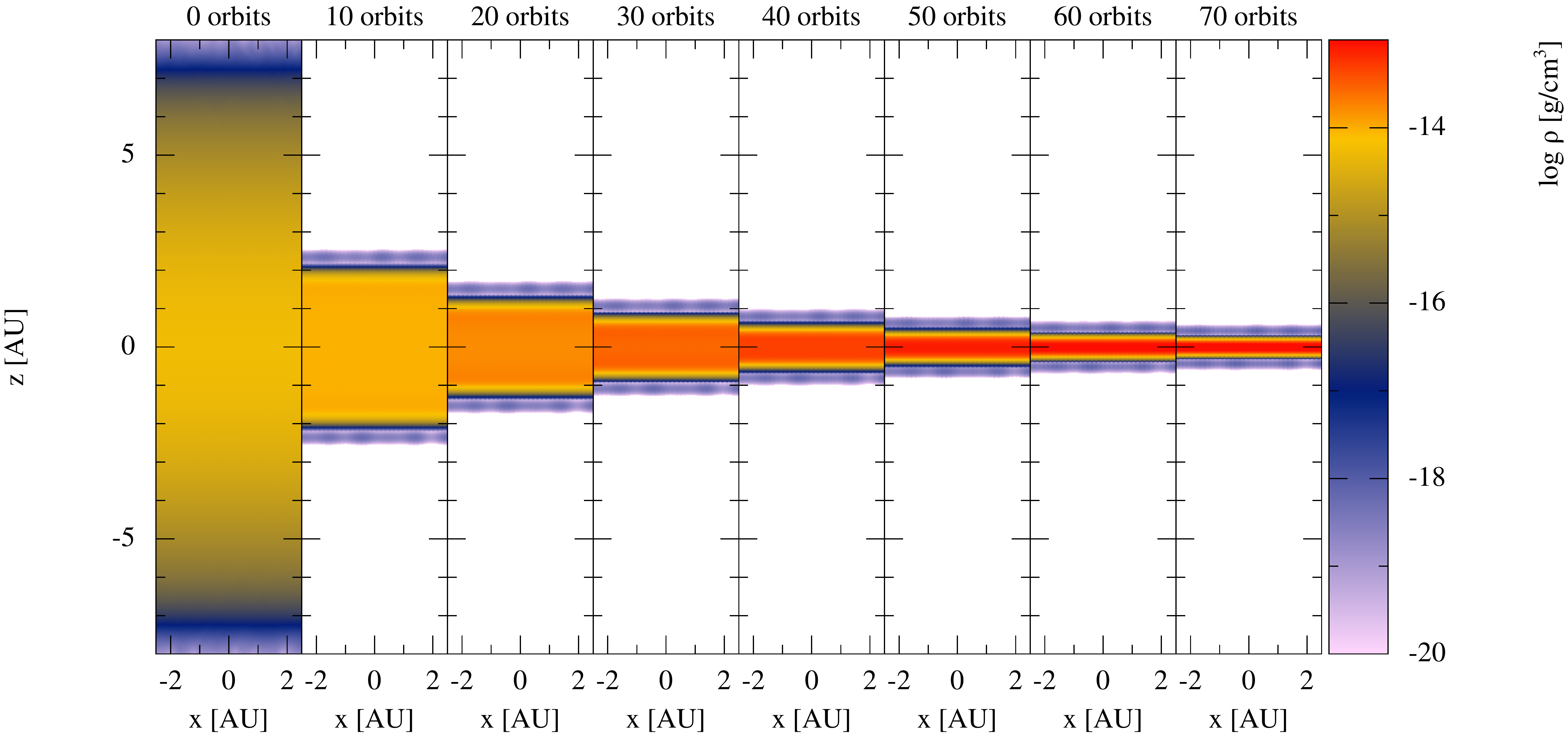}}
	\caption{Dust densities at $10$ orbit intervals for mm-sized dust grains in the \textsc{dustydisc} problem using $20\,100$ particles. Gravitationally, all particles are assumed to be located at $R=50\,$AU with a disc aspect ratio of $H/R=0.05$ and $H_0=2.5\,$AU.}
	\label{fig:disc_test_non-conserv_plot}
\end{figure*}

The \textsc{dustydisc} problem \citep{Price/Laibe/2015} is similar to the plane-parallel atmosphere described in \cref{sec:plane-parallel_atmosphere}. Its value lies in the fact that it tests the performance of our algorithms under physically relevant conditions. It probes how our unequal-mass, one-fluid formulation handles evolving a dust distribution concurrently in a wide spectrum of stopping times. This is different from previous tests which analysed each drag regime separately.

Anticipating the fact that we are forced to use unequal masses to reach photoevaporation densities, we restrict ourselves wholly to using unequal-mass particles from this point onwards. The particle masses for this test are initially assigned using the exponential distribution in \cref{eq:isothermal_density_profile} such that the initial density profile can be built entirely from uniformly spaced particles. Then, to obtain the correct density scaling, the masses are iteratively scaled until $|\rho_0-\rho_\text{max}|/\rho_0$ falls below a given tolerance. 

Because we have no analytic solution for this test, we compare our results against \citet{Price/Laibe/2015}. We replicate their physical parameters by choosing $\rho_\text{g,0} = 6 \times 10^{-13}\, \text{g}\, \text{cm}^{-3}$, $R = 50\,$AU (where again $R$ is a fixed parameter for \emph{all} particles), and an aspect ratio $H/R = 0.05$. The scale height $H = 2.5\,$AU is then used to calculate the initial sound speed according to $c_\text{s} = H\Omega_\text{K}$, where $\Omega_\text{K} = \sqrt{\mathcal{G} M / R^3}$. For a solar mass star, the orbital period is $t_\text{orb} \equiv 2 \pi \Omega_\text{K} \approx 353\,$years. We begin our simulation by placing $20\,100$ particles in a uniform (staggered) lattice inside a Cartesian box, $(x,z) \in [-3H,3H]$, with periodic boundary conditions in $x$ and open boundaries in $z$. We relax the system for $\sim \! 14$ orbital periods using artificial viscosity and a damping factor on the velocity as prescribed in \citet{Price/Laibe/2015}. The simulation is then restarted with mm-sized grains added to the disc with the same initial density profile as the gas, but scaled by a gas-to-dust ratio of $100$. The dust is then allowed to evolve for $70$ orbital periods.

In performing this test, we found that the physical conditions in atmospheres of settled discs---in particular, tiny dust fractions and large differential velocities---give rise to a numerical instability in the one-fluid SPH equations. Two methods for dealing with this instability are described in \cref{sec:numerical_instability}. Following the first of these methods, \Cref{fig:disc_test_non-conserv_plot} shows the temporal evolution of the dust density for $20\,100$ settling dust grains. Our results match those from \citet{Price/Laibe/2015}, thus validating the modifications contained in \cref{sec:numerical_instability_method_1}. Further comparison with their work shows that using unequal masses effectively extends our density range to much lower values than were previously possible.

\section{Numerical instability}
\label{sec:numerical_instability}

We found that using the conservative one-fluid SPH equations results in a numerical instability that causes $\epsilon$ and $\Delta \mathbf{v}$ to diverge. The instability only occurs where $\epsilon$ is small and $\Delta \mathbf{v}$ is large (e.g. the upper atmosphere of settled discs or photoevaporative winds with large dust grains). It cannot always be mitigated by using the non-conservative dissipation terms from \citetalias{Laibe/Price/2014b}. The same problem occurs when using equal-mass particles, but to a lesser degree because the resolution is much lower in the malignant areas of the disc where the instability occurs. Thus far, we have only found two ways to circumvent the instability.

\subsection{Method 1}
\label{sec:numerical_instability_method_1}

In the first method, we make three modifications to \cref{eq:onefluid_density,eq:onefluid_dustfrac,eq:onefluid_momentum,eq:onefluid_deltav,eq:onefluid_energy} that prevent the instability from occurring: (i) we use the additive form of the SPH gradient for the $\Delta v^2$ term in \cref{eq:SPH_deltav}, (ii) we use the non-conservative form of the $\Delta \mathbf{v}$ dissipation term, and (iii) we use the positive-definite formulation for the evolution of $\epsilon$. We discuss the reasons for each of these changes below:

(i) The term responsible for for the instability appears to be the penultimate term in \cref{eq:SPH_deltav}. Thus far, we have been unable to develop a dissipation term that is universally capable of controlling the growth of the instability. Using the additive form of the SPH gradient for this term is significantly more stable:
\begin{align}
	& \left( \frac{\text{d} \Delta \mathbf{v}}{\text{d} t}  \right)_\text{term}  = \frac{\rho_a}{2} \sum_b m_b 
	\left[ \frac{(1 - 2 \epsilon_a) \Delta \mathbf{v}_a^2}{\chi^a_b \rho_a^2} \nabla_a W_{ab}(h_a) \right. \nonumber
\\
	& \left. \phantom{ \left( \frac{\text{d} \Delta \mathbf{v}}{\text{d} t}  \right)_\text{term}  = \frac{\rho_a}{2} \sum_b m_b}
	+ \frac{(1 - 2 \epsilon_b) \Delta \mathbf{v}_b^2}{\chi^b_a \rho_b^2} \nabla_a W_{ab}(h_b) \right].
\end{align}
There is nothing inherently wrong about using this form of the SPH gradient; in fact, this same form is used for the pressure gradient in \cref{eq:SPH_momentum} and is second-order accurate. However, conservation laws provide constraints on the form of the SPH equations. Violating these rules results in a code that no longer conserves one or more of these properties.

(ii) We single out the dissipation term for $\Delta \mathbf{v}$ because it is the only term to contain the factor $\epsilon_b/\epsilon_a$. In most of the tests, this factor is approximately unity. However, as dust settles to the midplane in the \textsc{dustydisc} problem a large dispersion in $\epsilon$ develops in the atmosphere of the disc where the dust fraction becomes very small. If $\epsilon_a \ll \epsilon_b$ this term diverges and causes the code to fail. Removing this factor keeps this term from diverging, but again violates conservation properties.

(iii) We find that $\frac{\text{d} \epsilon}{\text{d} t}$ is very sensitive to small changes when $\epsilon$ is small. As a result, a large dispersion in $\epsilon$ develops with dust fractions oscillating wildly between positive and negative values. These oscillations are somewhat mitigated by using the positive-definite formulation for the dust fraction presented in \citet{Price/Laibe/2015} that evolves $S \equiv \sqrt{\rho \epsilon}$ (not to be confused with grain size) instead of $\epsilon$:

\begin{align}
	& \frac{\text{d} S}{\text{d} t}  = -\frac{\rho_a}{2} \sum_b m_b S_b \left( \frac{1-S_a^2/\rho_a}{\chi^a_b \rho_a^2} \Delta \mathbf{v}_a \cdot \nabla_a W_{ab}(h_a) \right. \nonumber
	\\	& \phantom{\frac{\text{d} S}{\text{d} t}  = -\frac{\rho_a}{2} \sum_b m_b S_b} \left. {} + \frac{1-S_b^2/\rho_b}{\chi^b_a \rho_b^2} \Delta \mathbf{v}_b \cdot \nabla_a W_{ab}(h_b)  \right) \nonumber
	\\	& \phantom{\frac{\text{d} S}{\text{d} t}  =} {} +  \sum_b m_b \frac{S_a}{2 \chi^a_b \rho_a} \mathbf{v}_{ab} \cdot \nabla_a W_{ab}(h_a),
\end{align}
Thus, despite giving slightly poorer results in other tests, using $S$ seems to give better results in regions where the dust fraction is small.

\subsection{Method 2}
\label{sec:numerical_instability_method_2}

The second method is much less involved. We find that the oscillations created by the instability can effectively be controlled by performing an artificial cut in the dust fraction at $\epsilon \sim 10^{-5}$. There is some risk of tainting the results when using this method, but the risk can be made very low by only employing the cut when the instability becomes unbounded. Small oscillations develop in $\epsilon$ whenever dust grains decouple from the gas and $\rho_\text{d} \to 0 $; however, the majority of these cases are benign and can easily be distinguished from physical dust by their low density, periodic structure, and physical context. When unbounded growth in $\epsilon$ does occur, it is always in regions where dust has clearly been evacuated, suggesting that there is little danger in performing the cut.\\

Neither method is inherently more accurate than the other. Thus, when forced to deal with the instability, we adopt Method 1 for stand-alone simulations---such as the \textsc{dustydisc} and \textsc{dustyphoto} tests---and adopt Method 2 when one or more simulations from a larger set goes unstable, to ensure that the fluid dynamics are consistently calculated across the entire set (like in \cref{sec:dusty_photoevaporation}). 


\bsp	
\label{lastpage}
\end{document}